\documentclass[fleqn,usenatbib]{mnras}
\usepackage{newtxtext,newtxmath}
\usepackage[T1]{fontenc}

\usepackage{graphicx}	\usepackage{amsmath}	\usepackage{xspace}
\usepackage{hyperref}

\newcommand{\msun}{\,{M$_{\odot}$}\xspace}
\newcommand{\ergs}{\,{erg\,s$^{-1}$}\xspace}
\newcommand{\sbunits}{\,erg\,s$^{-1}$\,cm$^{-2}$\,arcsec$^{-2}$\xspace}
\newcommand{\Lya}{\ifmmode{\mathrm{Ly}\alpha}\else Ly$\alpha$\xspace\fi}
\newcommand{\angstrom}{\text{\normalfont\AA}\xspace}
\newcommand{\ldens}{\ergs\,cMpc$^{-3}$\xspace}
\newcommand{\figpath}{./figures_opt}

\title[TNG50: the cosmic web in Lyman-alpha emission]{The cosmic web in Lyman-alpha emission}
\author[C. Byrohl and D. Nelson]{
Chris Byrohl,$^{1}$\thanks{E-mail: chris.byrohl@uni-heidelberg.de}
Dylan Nelson$^{1}$
\\
$^{1}$Universität Heidelberg, Zentrum für Astronomie, Institut für Theoretische Astrophysik, Albert-Ueberle-Str. 2, 69120 Heidelberg, Germany
}

\date{}
\pubyear{2022}

\begin{document}
\label{firstpage}
\pagerange{\pageref{firstpage}--\pageref{lastpage}}
\maketitle

\begin{abstract}
We develop a comprehensive theoretical model for Lyman-alpha emission, from the scale of individual Lyman-alpha emitters (LAEs) to Lyman-alpha halos (LAHs), Lyman-alpha blobs (LABs), and Lyman-alpha filaments (LAFs) of the diffuse cosmic web itself. To do so, we post-process the high-resolution TNG50 cosmological magnetohydrodynamical simulation with a Monte Carlo radiative transfer method to capture the resonant scattering process of Lyman-alpha photons. We build an emission model incorporating recombinations and collisions in diffuse gas, including radiative effects from nearby AGN, as well as emission sourced by stellar populations. Our treatment includes a physically motivated dust model, which we empirically calibrate to the observed LAE luminosity function. We then focus on the observability, and physical origin, of the $z=2$ Lyman-alpha cosmic web, studying the dominant emission mechanisms and spatial origins. We find that diffuse Lyman-alpha filaments are, in fact, illuminated by photons which originate, not from the intergalactic medium itself, but from \textit{within} galaxies and their gaseous halos. In our model, this emission is primarily sourced by intermediate mass halos ($10^{10} - 10^{11}$\msun), principally due to collisional excitations in their circumgalactic media as well as central, young stellar populations. Observationally, we make predictions for the abundance, area, linear size, and embedded halo/emitter populations within filaments. Adopting an isophotal surface brightness threshold of $10^{-20}$\sbunits, we predict a volume abundance of Lyman-alpha filaments of ${\sim}10^{-3}$\,cMpc$^{-3}$ for lengths above $400$\,pkpc. Given sufficiently large survey footprints, detection of the Lyman-alpha cosmic web is within reach of modern integral field spectrographs, including MUSE, VIRUS, and KCWI.
\end{abstract}


\section{Introduction}

Within the $\Lambda$CDM cosmological paradigm, gravitationally unstable initial matter density fluctuations evolve into a filament-dominated structure on large scales: the cosmic web~\citep{Bond96}. At late times, the majority of dark matter halos, as well as galaxies, reside in the filaments and nodes of this cosmic web~\citep{Meiksin09}. The same is true for the majority of dark matter and baryons, including diffuse gas. As a result, the formation and evolution of galaxies is mediated in large part by their gaseous environments, including gas gravitationally bound within dark matter halos -- the circumgalactic medium~\citep[CGM;][]{Tumlinson17}.

The large-scale filaments of the cosmic web can indirectly be observed through galaxy clustering in galaxy redshift surveys~\citep[e.g.][]{Colless01, Abazajian09}. At high redshift, a direct detection of these filaments is possible via absorption by the Lyman-alpha (\Lya) line of neutral hydrogen. In this case, spectra of background sources, mainly quasars, probe the hydrogen distribution in the intergalactic medium (IGM) along the line-of-sight~\citep{Gunn65, Meiksin09}. In recent years, high sampling density of such quasar spectra has enabled the reconstruction of the three-dimensional density field of neutral hydrogen~\citep{Lee14,Lee18,Newman20}. However, the coarse resolution of the order of megaparsecs makes it difficult to resolve the filamentary structure of the cosmic web, a limitation inherited from the sparseness of background quasars on the sky.

In contrast to absorption, the \Lya emitting cosmic web offers a complementary approach. However, direct imaging of large-scale \Lya filaments remains challenging given the low emissivities of the diffuse gas~\citep{Gallego18}. For denser environments, \Lya emission is already a frequently used tracer of cold gas. For example, \Lya emission is commonly used to identify high-redshift galaxies in blind surveys~\citep{Cowie98}. In targeted surveys, extended emission with sizes of ${\sim}10-100$\,pkpc around massive galaxies has been detected for decades~\citep{McCarthy87, Heckman91, Steidel00}. More recently, the \Lya emission around smaller star-forming galaxies, tracing the CGM around these objects, has been revealed on scales of ${\sim}10$\,pkpc --- first through narrowband stacking~\citep{Hayashino04, Steidel11, Matsuda12, Momose14, Kakuma21} and then through integral field spectroscopy~\citep{Wisotzki16, Leclercq17, LujanNiemeyer22a}.

The latest observations of \Lya emission around star-forming galaxies show flattened extended radial profiles~\citep{Wisotzki18, Kakuma21, Kikuchihara21, LujanNiemeyer22}, potentially hinting at the faint \Lya glow of the cosmic web. Large filamentary structures with extents of $\gtrsim 1$\,pMpc have been detected by targeting known overdense fields: the SSA22 protocluster \citep{Umehata19, Herenz20} and the Hyperion proto-supercluster \citep{Huang22}. Aligned stacking of galaxy pairs at $z \sim 2$ shows extended \Lya emission from the CGM, but no signal from intergalactic scales~\citep{Gallego18}. However, the \Lya cosmic web at $z\gtrsim 3$ has potentially recently been detected in a blind survey using integral field spectroscopy~\citep{Bacon21}.

The \Lya emission line of the neutral hydrogen atom is a promising tool to study large-scale structure, and can be probed at redshifts $z\geq 2$ with current ground-based instruments on ${\sim}8-10$\,m telescopes such as MUSE, KCWI and VIRUS~\citep{Bacon10, Morrissey18, Gebhardt21} on the VLT, Keck, HET telescopes respectively. Upcoming ${\sim}30$\,m class telescopes will further enable the detection of \Lya filaments. The lower end of the redshift range at $z\sim 2$ is particularly promising given its favorable cosmological surface brightness (SB) dimming scaling as ${(1+z)}^{-4}$. For example, the majority of filament candidates in~\citet{Bacon21} are located towards the lowest accessible redshifts, around $z\sim 3$. However, the overall redshift trend of filament detectability depends on a complex evolution of the physical properties of filaments, including their density, temperature, and ionization state, as well as galaxy clustering, global star-formation, dust content and IGM opacity.

Predictions for the observability of cosmic web filaments have been made for intensity mapping~\citep{Silva13, Silva16, Heneka17} as well as direct observation~\citep{Elias20, Witstok21}. However, this requires comprehensive and accurate emission models for \Lya photons. The physical processes involved include emission from excitations and recombinations in the diffuse gas, and the effective emission which arises due to the radiative output of young stars during the process of star formation, as well as due to radiation from AGN.

The uncertainties in the modeling of emission mechanisms are further complicated by the complex radiation transfer that \Lya photons experience~\citep{Neufeld90, Neufeld91, Hansen06}. The \Lya emission line is resonant and optically thick in astrophysical environments, leading to numerous scatterings before photons eventually escape, or are destroyed. This causes substantial spatial and spectral redistribution of photons, making this forward modeling step imperative from the simulation point of view, as well as complicating the interpretation of any observed emission.

A common approach is to post-process cosmological (radiation-) hydrodynamical simulations with Monte Carlo based \Lya radiative transfer codes~\citep[e.g.][]{Cantalupo05, Laursen07, Kollmeier10, Goerdt10} to study different \Lya observables such as LAE clustering~\citep{Zheng11, Behrens18, Byrohl19}, spectral signatures from the IGM~\citep{Byrohl20, Park22} and extended emission.

Many recent theoretical studies of extended \Lya emission have included scattering effects and focused on CGM scales~\citep{Lake15, Gronke17a, Behrens19, Smith19a, Mitchell21, Byrohl21}. However, investigations dedicated to the cosmic web in \Lya emission have universally neglected the impact of radiative transfer~\citep{Elias20, Witstok21}. In that context, \citet{Witstok21} conclude that observations of the cosmic web at lower redshifts are most promising in overdense regions, where emission is dominated by the halos and galaxies within filaments. In this regime, collisional excitations produce more \Lya photons than recombinations. Simultaneously, \citet{Elias20} suggest that \Lya surface brightness predictions can be used to constrain the underlying galaxy formation model physics. However, the lack of a quantitative assessment to date for the occurrence of these filaments hinders observational constraints on \Lya radiative transfer simulations of the cosmic web.

In this study, we model and characterize the Lyman-$\alpha$ cosmic web in emission. To do so, we adopt the high-resolution, large-volume TNG50 cosmological galaxy formation simulation. We focus on redshift $z=2$ and post-process the original simulation output with our sophisticated Monte Carlo radiative transfer method. We furthermore introduce a physically motivated dust rescaling model calibrated against the observed \Lya luminosity function (LF).

This paper is organized as follows: In Section~\ref{sec:methodology}, we introduce the TNG50 simulations, the \Lya radiative transfer code, the underlying emission model, and our analysis pipeline. In Section~\ref{sec:results}, we present results regarding global \Lya related properties and a study of filamentary \Lya structures. In Section~\ref{sec:discussion}, we discuss our results for the dominant physical mechanisms which light up the \Lya cosmic web, the origin of \Lya photons from filaments, and the detectability of the cosmic web with current and upcoming \Lya emission surveys.

\section{Methodology}
\label{sec:methodology}

\subsection{TNG50}

The TNG50 simulation~\citep{Pillepich19,Nelson19a} is the highest-resolution simulation of the IllustrisTNG suite, a series of three large-volume magnetohydrodynamical cosmological simulations~\citep{Pillepich18,Naiman18,Nelson18,Marinacci18,Springel18}. The simulations were run with the {\textsc AREPO} code~\citep{Springel10}, which solves the coupled equations for self-gravity and ideal, continuum magnetohydrodynamics~\citep{Pakmor11} discretizing space using an unstructured Voronoi tessellation. The TNG galaxy formation model~\citep{Weinberger17, Pillepich18a} includes a treatment for the majority of physical processes shaping galaxy formation: primordial and metal-line cooling, heating from ultraviolet background (UVB) radiation, star formation above a density threshold, stellar feedback driven galactic winds, stellar population evolution and chemical enrichment from supernovae Ia, II and AGB stars, and the seeding, merging, and growth via accretion of supermassive black holes (SMBHs).

The temperature and ionization state of gas, which is relevant for \Lya emission and scattering, is computed following the primordial cooling network of~\citet{Katz96} with additional metal-line cooling from \textsc{CLOUDY} cooling tables. In addition, a heating and ionization term arises from the assumption of a uniform, time-varying UVB using the intensities given in~\citet[][FG11 update]{Faucher-Giguere09}. An additional local ionization field is introduced for active galactic nuclei (AGN) up to $3$ times the hosting halos' virial radius. The underlying AGN luminosities are proportional to the accretion rates above a certain accretion threshold, modulated by an obscuration factor based on~\citet{Hopkins07}, assuming optically thin gas~\citep{Vogelsberger13}. Radiation from the UVB and AGN is attenuated on-the-fly according to~\citet{Rahmati13} to account for self-shielding. The additional AGN radiation field is important for the gas state and subsequent \Lya emission. Particularly at $z\geq 2$ substantial changes arise given the high accretion rates of SMBHs~\citep[see][]{Byrohl21}. Note that ionizing radiation escaping from \textit{local} stellar sources is not included in the model.

TNG50 has a gas mass resolution of $m_{\rm bayron} = 8.5 \times 10^4$\msun and a dark matter mass resolution of $m_{\rm DM} = 4.5 \times 10^5$\msun. This roughly corresponds to a spatial resolution of ${\sim}100$\,physical pc in the ISM\@. The simulations use a set of cosmological parameters consistent with recent results by the Planck collaboration \citep{PlanckCollaboration16}, namely $\Omega_{\Lambda,0}=0.6911$, $\Omega_{m,0}=0.3089$, $\Omega_{b,0}=0.0486$, $\sigma_8=0.8159$, $n_s=0.9667$ and $h=0.6774$.

\subsection{Lyman-alpha emission and radiative transfer}
\label{sec:emission}

We calculate the \Lya radiative transfer in post-processing using our updated, light-weight line emission radiative transfer code~\citep[originally introduced in][]{Behrens19}. It propagates Monte Carlo photon packages according to the given gas structure, accounting for scattering and destruction. Upon each scattering, we calculate the luminosity contribution escaping towards predefined observers~\citep[the ``peeling-off'' algorithm;][]{Whitney11}. Our radiative transfer method supports a range of geometries, including the underlying Voronoi tessellation of TNG50, and for this work we compute and ray-trace through the global (entire snapshot) mesh at once, in order to self-consistently capture environmental and large-scale IGM effects in the radiative transfer~\citep[as introduced in][]{Byrohl21}.

We follow the \Lya emission model as introduced in~\citet{Byrohl21}, where we include the emission of diffuse gas by recombinations and collisional excitations, and emission from star-forming regions. The emission model for the diffuse gas is unchanged, with luminosity densities given by
\begin{equation} \label{eq:lumrec}
\epsilon_\mathrm{rec} = f_\mathrm{rec}(T) \,n_\mathrm{e} \,n_\mathrm{HII} \,\alpha(T) \,E_{\mathrm{Ly}\alpha}
\end{equation}
and
\begin{equation} \label{eq:lumexc}
\epsilon_\mathrm{coll} = \gamma_{\mathrm{1s2p}}(T) \,n_\mathrm{e} \,n_\mathrm{HI} \,E_{\mathrm{Ly}\alpha},
\end{equation}
which scale with the number density of electrons ($n_\mathrm{e}$), neutral ($n_\mathrm{HI}$) and ionized hydrogen ($n_\mathrm{HII}$). The temperature dependent recombination ($\alpha$) and collisional excitation coefficient $\gamma_\mathrm{1s2p}(T)$ are taken from~\citep{Scholz90,Scholz91,Draine11}.

For the emission from dense gas around star-forming regions, we update the previous model and do not emit \Lya radiation based on the instantaneous star formation rate of gas cells. Instead, we model the \Lya emission from stellar populations as follows. We calculate the ionization rate $\dot{N}_{\mathrm{ion}}$ of all stars according to their age, mass, and metallicity with \textsc{BPASS}~\citep{Stanway18} assuming a Chabrier initial mass function~\citep{Chabrier03}. From this, we derive the nebular emission of \Lya under the case-B assumption via
\begin{align}
\label{eq:LyaSF}
  L_{\Lya, \mathrm{orig}} = f_{B}\cdot\left(1-f_{\mathrm{UV,esc}}\right)\cdot E_{\Lya}\cdot \dot{N}_{\mathrm{ion}}.
\end{align}
This approach becomes feasible given the high resolution of TNG50, which allows a reasonable sampling of young $< 10$\,Myr stellar populations, a common problem for cosmological simulations at lower resolution~\citep{Trayford17, Nelson18}.

In addition to the ionization rate $\dot{N}_{\mathrm{ion}}$, we calculate the luminosity per wavelength of the stellar continuum $l_{\mathrm{cont}}$ at the Lyman-$\alpha$ resonance wavelength with a Gaussian smoothing kernel of $\sigma=50$\angstrom. We then compute the intrinsic rest-frame equivalent width (REW) as
\begin{align}
\label{eq:REW}
    \mathrm{REW} = L_\Lya/l_\mathrm{cont}.
\end{align}

For the diffuse emission mechanisms, we spawn one photon per gas cell. For stellar emission, we spawn 11000 photons per $10^{42}$\ergs in luminosity, with a minimum of 3 photons per stellar particle.

In the radiative transfer code, we use the temperature and neutral hydrogen density as directly inferred from the TNG50 snapshot data. The effective temperature and average hydrogen density cannot be used in a straightforward way in star-forming cells invoking a sub-grid effective equation of state for a two-phase ISM~\citep{Springel03}. For these cells, we adopt the temperature and density values from the snapshot, but do not generate any emission from recombinations and excitations.

The role of dust for \Lya radiative transfer is important~\citep{Laursen09, Byrohl21}. Its impact is particularly susceptible to the unresolved small-scale structure~\citep{Gronke17} due to the high optical depths within individual cells. We have developed new models and strategies for incorporating dust, and for the present work have decided on an empirical calibration strategy. In particular, we introduce an effective dust attenuation by rescaling the stellar luminosity contributions as described in Section~\ref{sec:rescaling}.

All photons are emitted at the \Lya line-center frequency with a random initial direction. Frequency shifts with $\leq 200$\,km$\,$s$^{-1}$ injected on the ISM scale have little impact on the radiative transfer for the described setup~\citep[see appendix in][]{Byrohl21}\footnote{The impact of the frequency shift depends on the simulated density and velocity structure on ISM and CGM scales. Strictly speaking, these results therefore only hold within the TNG50 simulation.}. Larger wavelength shifts could however largely change the outcome. Generally speaking, a large spectral redshift (blueshift) decreases (increases) the redistribution of photons into their surroundings.

\subsection{Scattered Lyman-alpha photon properties}

\begin{figure*}
\centering
\includegraphics{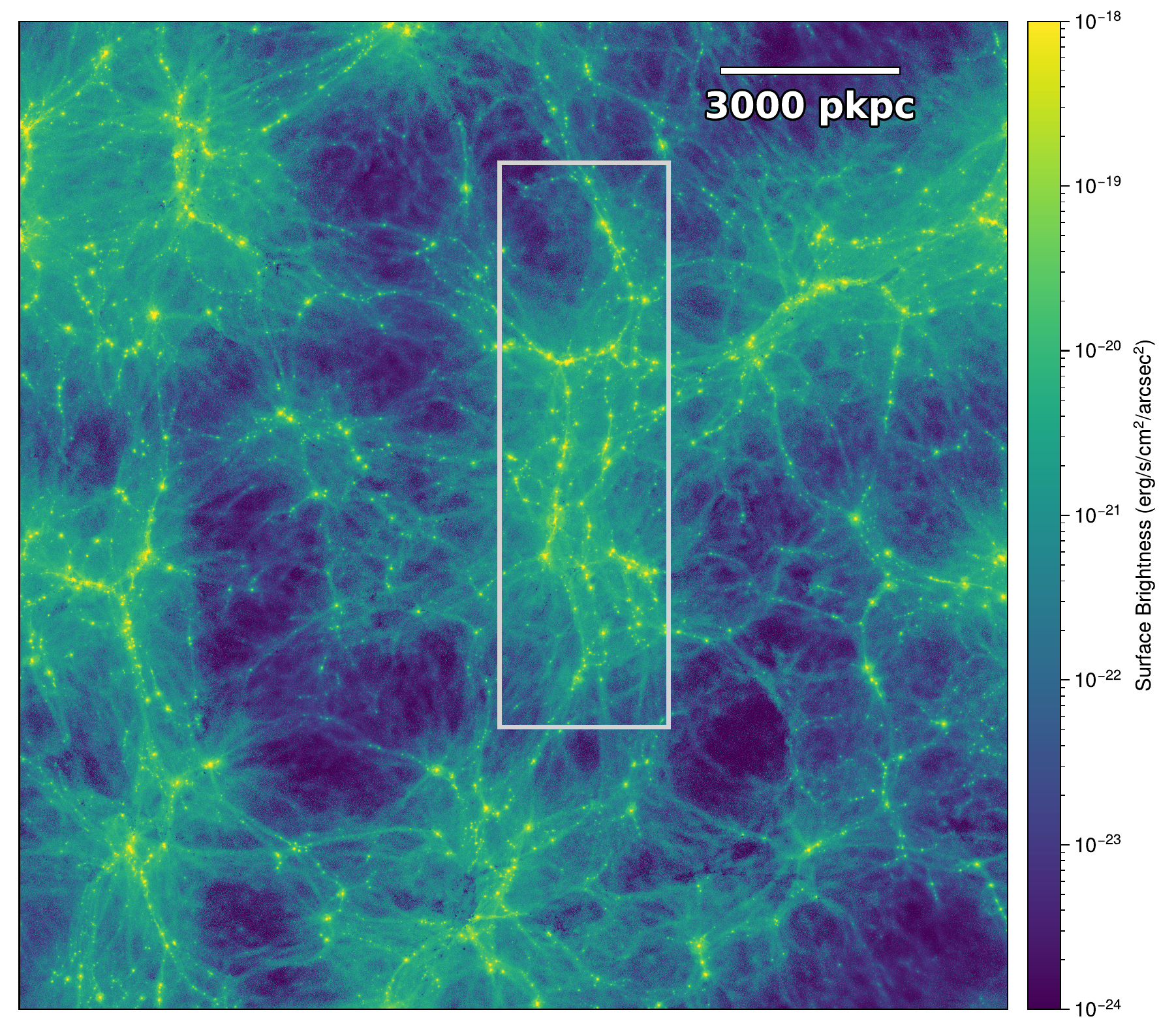}
 \caption{Overview visualization of \Lya surface brightness with a slice depth of $5.7$\angstrom in the observed frame for TNG50-1 at $z=2$. \Lya emission traces the large-scale cosmic web with contiguous filamentary structures up to ${\sim}10$\,pMpc for surface brightnesses above $10^{-21}$\sbunits. Knots above $10^{-18}$\sbunits show potential \Lya blobs. The white rectangle marks a zoom-in region that we focus on in the following analysis.}
\label{fig:overviewplot}
\end{figure*}

For each photon contribution, we keep track of the following details: the spatial location of last scattering, luminosity, and wavelength. In addition, we save the global Voronoi cell indices at the points of initial emission and last scattering. This allows us to trace and analyze the photon contributions with respect to the underlying simulation, including its halo and galaxy populations and their properties.

We distinguish between two different sets of photons:
\begin{enumerate}
 \item \textbf{intrinsic}: \Lya photons as emitted from gas cells and stellar particles, intentionally neglecting further gas interaction, i.e.\ scatterings and destruction.
 \item \textbf{scattered}: \Lya photons which escape toward the observer after each scattering of a previously emitted \Lya photon, including attenuation and scattering.
\end{enumerate}

Only the scattered photons represent observable \Lya signatures. However, the intrinsic photons, which are only accessible theoretically, enable us to study the origin of \Lya emission and the impact of the \Lya radiative transfer.

By identifying the initial and final Voronoi gas cell for each photon, we classify each photon into exactly one of the following five spatial categories, at the time of emission as well as last scattering:

\begin{enumerate}
 \item \textbf{IGM}: intergalactic medium gas, i.e.\ does not belong to any collapsed halo.
 \item \textbf{outer halo}: gas which is part of a dark matter halo, but gravitationally unbound, i.e.\ on the outskirts.
 \item \textbf{CGM}: gas in the halo, gravitationally bound to the central galaxy, and outside 10\% of the halo virial radius.
 \item \textbf{central}: gas in the halo, gravitationally bound to the central galaxy, and inside 10\% of the halo virial radius.
 \item \textbf{satellite}: gas gravitationally bound to a satellite galaxy which is within a larger host halo.
\end{enumerate}

These categories rely on the Friends-of-Friends halo and \textsc{subfind} subhalo identification algorithms~\cite[see][]{Nelson19}.

Furthermore, we can study the physical gas state, as well as galaxy and halo properties, at two distinct times:

\begin{enumerate}
    \item \textbf{at origin}, using the gas cell where the \Lya photon was initially emitted, or
    \item \textbf{at last scattering}, using the gas cell within which the \Lya photon finally escaped to the observer.
\end{enumerate}

We create 2D surface brightness projections with a depth of $5.7$\angstrom in the observed frame ($R\sim 650$), similar to the HET-VIRUS resolution~\citep{Hill21} and $50\%$ higher than the spectral binning used in~\citet{Bacon21} to study the \Lya cosmic web at higher redshift.

For surface brightness maps covering the entire extent of the simulation box, we use a map with $4096^{2}$ pixels corresponding to resolution elements of ${\sim}0.5^{2}$\,arcsec$^2$. This resolution is also used for all reduced statistics, such as filament sizes and shapes. In such cases, we create and use as many projections of the given depth as possible, equally spaced and non-overlapping along the line-of-sight. When zooming into sub-regions, we use resolution elements of $0.2^{2}$\,arcsec$^{2}$.

\subsection{Observational calibration}
\label{sec:rescaling}

The escape or destruction of \Lya emission in the ISM strongly depends on small-scale gas structure. The relevant scales are at least partially unresolved at the resolution of TNG50, motivating us to develop a sub-grid attenuation model which is empirically calibrated. Instead of explicitly modeling the abundance, distribution and physics of dust, we instead rescale the \Lya luminosities in the ISM\@. Given the significantly lower optical depths of dust in more diffuse gas, we do not rescale the contributions from recombinations and collisional excitations, which occur only in non star-forming gas.

The rescaling of emission arising from nebular emission around stellar populations is done as follows. We first calculate a \Lya luminosity for each galaxy, by summing the luminosities of all gas and stars bound to the halo, restricted to a circular aperture with a diameter of 3 arcseconds. This procedure is done for the scattered photons of a full radiative transfer run, i.e. fully observable \Lya photons. Next, all photon contributions from the `stellar origin' are rescaled downward to match observational constraints. In particular, we use the $z=2$ \Lya luminosity function as our only calibration.

We propose a coarse but physically motivated attenuation model intended to represent \Lya destruction by dust.
We rescale the original luminosity of each stellar population $L_{\mathrm{orig},i}$ as
\begin{align}
  L_{\mathrm{int,i}} = f_{j}\,L_{\mathrm{orig,i}}
\end{align}
where $f_{j}$ is the per-halo rescaling factor for halo $j$,
\begin{align}
f_{j}=\exp\left[-\tau_{j}\right]
\end{align}
that is set by the attenuating optical depth $\tau_{j}$ for that particular halo. These optical depths are assumed to follow a simple relation with host halo properties, including scatter. In particular, they are drawn from a Gaussian with mean $\mu_{\tau}$ and standard deviation $\sigma_{\tau}$, similar in spirit to~\citet{Inoue18}. We parameterize the mean value using the mass-weighted average gas metallicity $Z_{j}$ of the host halo:
\begin{align}
\label{eq:rescaling}
 \mu_{\tau} = {\left(\frac{Z_{j}}{Z_{\star}}\right)}^{\beta}
\end{align}
and assume the scatter increases with the mean $\sigma_{\tau}= s \mu_{\tau}$. The three free fit parameters of our model are therefore: $Z_{\star}$, $\beta$ and $s$. Here, $\tau_{j}$ can be interpreted as the effective optical depth experienced by the Lyman-$\alpha$ photons. However, it can also capture potential disagreements of TNG50's star-formation rates with observations, compensate for modeling deficiencies of the diffuse emission, and encapsulate the impact of  different $L_{\Lya}\left(\dot{N}_{\mathrm{ion}}\right)$ relations.

We minimize the mean-squared error between the mocked TNG50 \Lya luminosity function and the observational data by~\citet{Konno16}. We impose a rest-frame equivalent width cut of $20$\angstrom, which is most appropriate for comparison with~\citet{Konno16} and only fit observational data points between $10^{42}$ and $10^{43}$\ergs. At $z=2$, the global best/fit model occupies a well-defined minimum with  $Z_{\star}=10^{-4.78}$, $\beta=0.47$ and $s=0.32$.

The intrinsic equivalent widths are modeled using the stellar continuum estimate as described in Section~\ref{sec:emission} attenuated by dust. For the dust attenuation of the continuum, we compute the optical depth
\begin{align}
\label{eq:uvdust}
\tau_\lambda^{\mathrm{dust}} = \left(\frac{A_\lambda}{A_V}\right) (Z_g/Z_\odot) (N_H/N_{H,0})
\end{align}
for dust similar to~\citet{Nelson18}, taking attenuation strictly proportional to the gas metallicity $Z_{g}$ and hydrogen column density $N_{H}$ with $Z_{\odot}=0.0127$ and $N_{H,0}=2.1 \cdot 10^{21}$\,cm$^{-2}$. We adopt the attenuation curve $\left(\frac{A_\lambda}{A_V}\right)$ from~\citet{Calzetti00}. The optical depth is calculated along each line-of-sight by ray-tracing through the metallicity and hydrogen density in each Voronoi cell.

In all cases, we use the rescaled luminosities throughout the following analysis by rescaling all photons, whether scattered or intrinsic, originating from stellar populations in halo $j$ by $f_{j}$ provided by our best fit model. Overall, this calibration ensures that our \Lya emission model is reasonable, and so fulfills a necessary condition to study the observability of the cosmic web hosting the \Lya emitting objects contained in the luminosity function\@.

\section{Results}
\label{sec:results}

We introduce the outcome of our theoretical modeling with Figure~\ref{fig:overviewplot}, which shows the \Lya surface brightness across cosmological scales for TNG50 at $z=2$. We project through a relatively narrow slice depth of $5.7$\angstrom, and adopt our fiducial emission model. The volume is suffused with \Lya light across a range of scales, from compact emission sources to elongated filamentary structures spanning megaparsecs to tens of megaparsecs in extent. Surface brightness levels within these filaments vary significantly, with brighter structures exceeding $10^{-20}$\sbunits.

\subsection{The Lyman-alpha luminosity function, and relation to galaxy and halo properties}
\label{sec:lfs}

\begin{figure}
\centering
\includegraphics[width=0.49\textwidth]{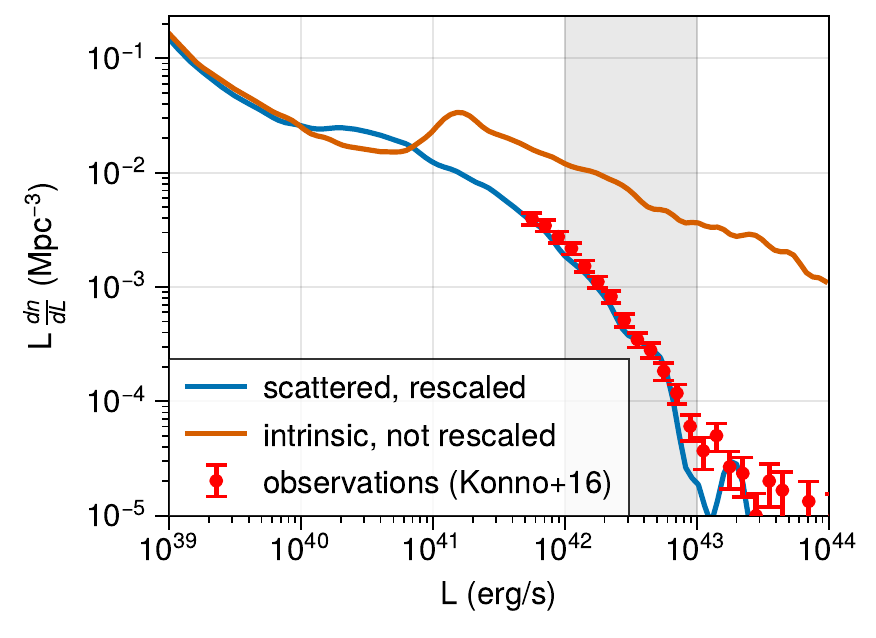}
 \caption{\Lya luminosity function of TNG50 at $z=2$. We compare the LF from scattered, rescaled luminosities (blue) and the intrinsic luminosities before calibration (orange). The observational data used in the calibration from \protect\citet{Konno16} are shown as red points with error bars. The gray band marks the observational data points between $10^{42}$ and $10^{43}$\ergs used in the fitting procedure. The luminosity function is the only observable statistic used in the calibration. Our simple model to account for the stochastic impact of dust is flexible enough to fit the observational data. The calibration largely dominates over radiative transfer in the luminosity function, such that the line for the scattered, not rescaled luminosity function would be close to the intrinsic, not rescaled (orange) line.}
\label{fig:lf}
\end{figure}

The realism of our \Lya modeling results bears close inspection. We first assess the outcome by considering the luminosity function (LF), the observable against which we calibrate the emission model. In Figure~\ref{fig:lf} we show the LF for the calibrated luminosities from scattered photons (blue) and for the intrinsic uncalibrated luminosities (orange). The calibrated (i.e. ``rescaled'', two terms we use interchangeably) luminosity function is in good agreement with the observational LF of~\citet{Konno16}. At higher luminosities $L> 10^{43}$\ergs where AGN contamination in the observed sample increases~\citep{deLaVieuville19} and the volume of TNG50 is too small to include these rare systems, we no longer (aim to) match observational data points. At luminosities below $10^{42}$\ergs, the calibrated TNG50 luminosity function gradually flattens with a plateau at ${\sim}10^{41}$\ergs before once more steepening, and then finally turning over and decreasing below ${\sim}10^{37}$\ergs (not shown). The plateau roughly coincides with star-formation ceasing to be the dominant emission mechanism in this luminosity range.

The global \Lya luminosity density inferred from the luminosity function integrated for all $L>10^{41.75}$\ergs, i.e.\ where constrained by the data points, is $5.2\cdot 10^{39}$\ergs\,cMpc$^{-3}$, which is roughly a factor of two smaller than the \citet{Konno16} data itself. This is due to the simulation's drop-off at the high-luminosity end compared to the distinct AGN induced bump in observations. We also point out that the majority of the global \Lya luminosity density is below the \citet{Konno16} lower limit of $L_{\min}=10^{41.75}$\ergs with a total luminosity density from the luminosity function of $1.33 \cdot 10^{40}$\ergs\,cMpc$^{-3}$, when we also neglect any REW threshold.

\begin{figure}
\centering
\includegraphics[width=0.49\textwidth]{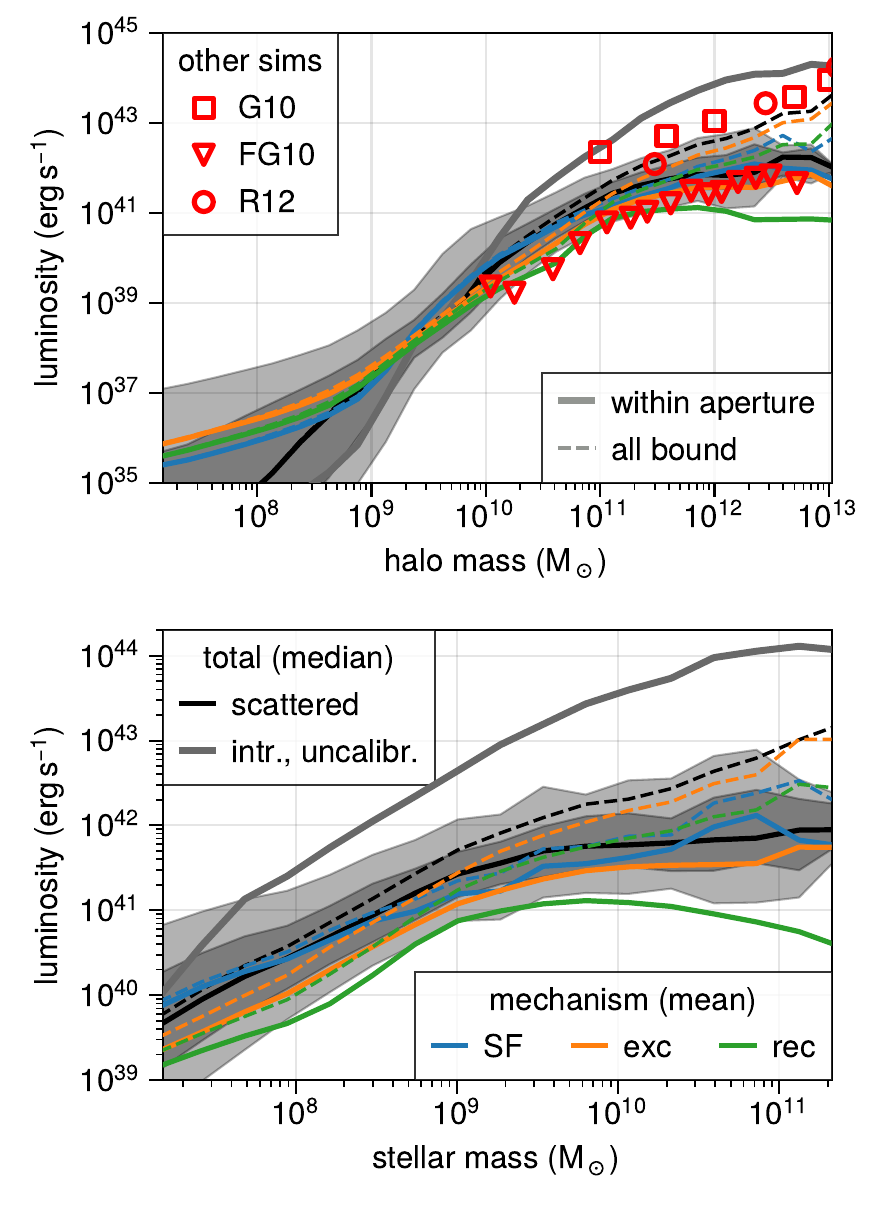}
 \caption{\Lya luminosity as a function of halo mass (top) and stellar mass (bottom). The median luminosity for scattered photons in our fiducial, observationally calibrated model is shown in black, with gray shaded regions showing the central $68$\% and $95$\% halo-to-halo variation. We also show the intrinsic, uncalibrated photon luminosity in light gray. Overall, \Lya luminosity rapidly and monotonically increases with mass. The relation between luminosity and halo mass is steepest between $10^{9}$ and $10^{10}$\msun, flattening at both lower and higher masses. The \Lya luminosities of galaxies and halos are (i) always lower for our fiducial model compared to the intrinsic (and uncalibrated) emission, and (ii) flatten faster due to increasing dust content. In addition, different line colors show the mean luminosity decomposed into contributions from stellar populations (blue), excitation (orange), and recombinations (green). For solid lines, the luminosity is computed following our fiducial definition (see text), while dashed lines show the luminosity for all bound gas without restriction by any aperture radius. For comparison, red symbols show selected computational results at $z=3$~\citep[G10, FG10, R10, ][]{Goerdt10, Faucher-Giguere10, Rosdahl12}. Our fiducial observable-mass relation is roughly bracketed by this range of previous models. The black curves represent the fundamental scaling relations for \Lya emission, and reflect the combination of the underlying TNG50 hydrodynamical simulation with our emission and radiative transfer model.}
\label{fig:halolums}
\end{figure}

In Figure~\ref{fig:halolums}, we show a fundamental outcome of the model: the \Lya mass-observable relations. Specifically, median \Lya luminosity as a function of halo mass (top panel) and stellar mass (bottom panel). The upper gray line shows the median of the total, intrinsic, uncalibrated luminosities. All other lines show the luminosity of our fiducial model after the calibration method and radiative transfer calculation. In particular, the black line shows the median of the total, scattered calibrated luminosities. Shaded regions show the central $68$\% and $95$\% of luminosities for scattered photons. Colored lines show the mean luminosities, for scattered photons, separating into the three physical origins: star formation i.e. young stellar population sourced (blue), collisions (orange), and recombinations (green).

All solid lines adopt our fiducial aperture and definition: summing photons from bound gas within an aperture radius of $1.5$\,arcsec. The dashed lines include contributions outside this aperture radius. The virial mass corresponding to this aperture radius is $5\cdot 10^{9}$\msun, above which the dashed lines rise above the solid lines in the upper panel. We find that for the most massive halos, the majority of escaping \Lya photons originate from radii beyond this aperture.

Overall, we see that \Lya luminosity monotonically increases with mass. The steepest scaling between halo mass and \Lya luminosity occurs between $10^{9}$ and $10^{10}$\msun. At lower masses, the relation flattens out as scattered photons from other halos start to dominate. At higher masses, the relation flattens significantly, in part due to the halo extent exceeding the fiducial aperture radius (compare to dashed lines), but primarily due to the impact of dust. \Lya luminosity after radiative transfer and calibration decreases substantially compared to intrinsic uncalibrated values, up to $2$\,dex for the most massive halos.

Our \Lya luminosity versus mass relations from Figure~\ref{fig:halolums} can be contrasted against other simulations, providing a benchmark comparison. As with other mass-observable relations, they can in the future also be compared against observational data, when large non-\Lya selected surveys are available.

To provide a first comparison, red markers show results from selected computational models at $z=3$~\citep{Goerdt10, Faucher-Giguere10, Rosdahl12}. All data points include the luminosity for all bound halo gas, and thus should be compared against the black dashed lines. Given substantial physical model and simulation differences, including the details of stellar and AGN feedback, cooling, and on-the-fly self-shielding in TNG with respect to these previous simulations, differences are to be expected. We discuss this in more detail in Section~\ref{sec:discussem}. Nevertheless, our TNG50 results are roughly bracketed from below by the~\cite{Faucher-Giguere10}\footnote{We compare against their self-shielding model ``\#9'', which corresponds to our model excluding diffuse emission from star-forming multiphase gas.} result, and~\cite{Goerdt10, Rosdahl12} from above. These studies focus on emission from collisional excitations and recombinations\footnote{Recombinations are neglected in~\citet{Goerdt10}.} without an explicit treatment for star-formation, and with exception of~\citet{Faucher-Giguere10}, without considering \Lya scattering.

\begin{figure}
\centering
\includegraphics[width=0.47\textwidth]{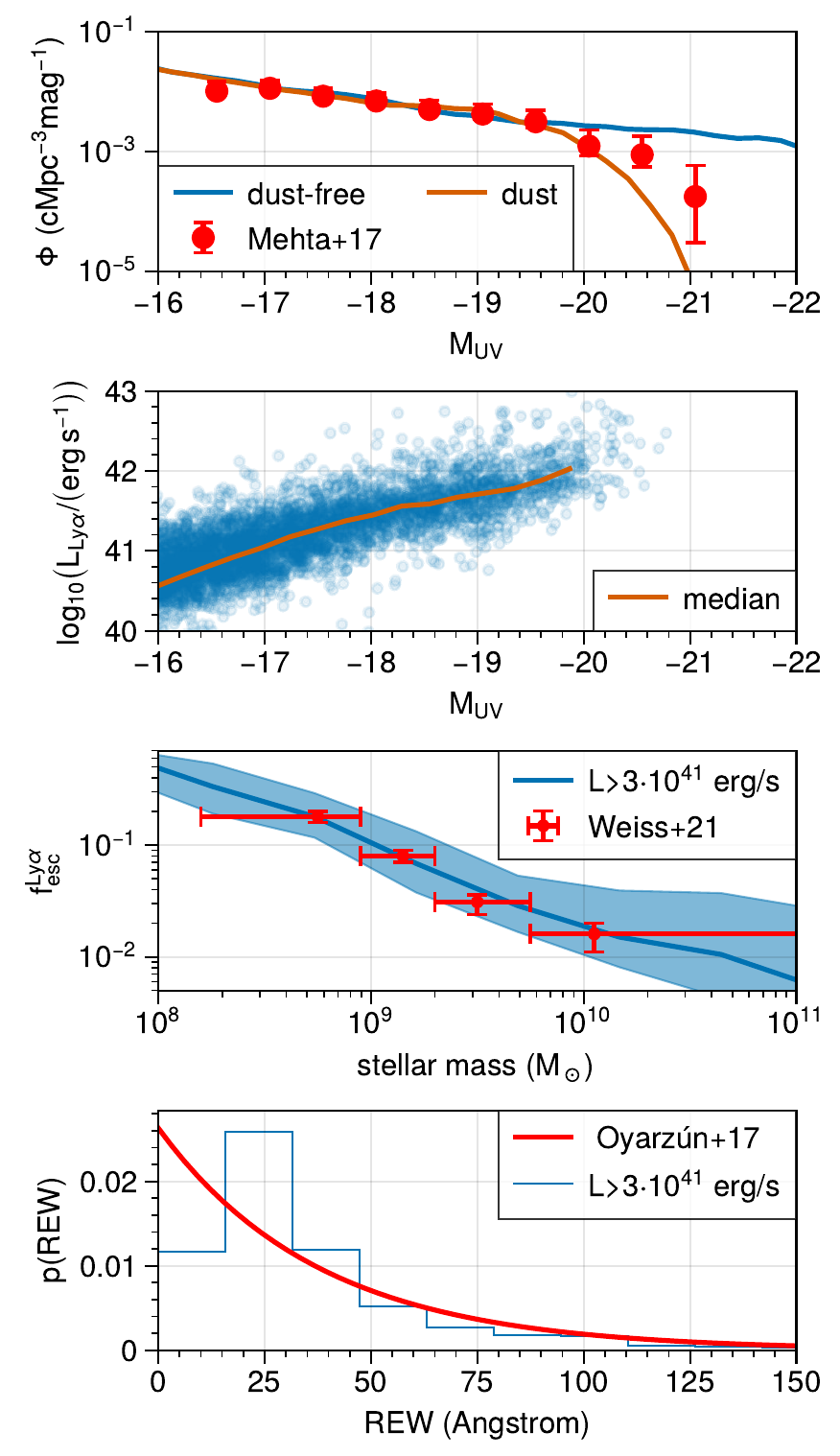}
  \caption{Observable properties and scaling relations of the \Lya emitting galaxy population at $z=2$ in contrast to data (where available). As none of these observations were used during calibration, this serves as a validation of our rescaling model. The first panel shows the UV luminosity function: the intrinsic (i.e.\ dust-free) simulation result in blue is too high at the high-mass end, while the calibrated LF (i.e.\ including the impact of dust, in orange) is a better fit to the data. The second panel shows the \Lya luminosity for scattered photons as a function of UV luminosity. The third panel plots \Lya escape fraction f$_{\mathrm{esc},\Lya}$ (see text for details) as a function of stellar mass, using the same aperture as for the \Lya and UV luminosities. In the last panel, we show the distribution of rest-frame equivalent widths. The last two panels impose a lower \Lya luminosity limit of $3\cdot 10^{41}$\ergs to best compare against the data. In red, we show observations from~\citet{Oyarzun17, Mehta17, Weiss21}. Our model reproduces these key properties of \Lya emitting galaxies in reasonable agreement with data, suggesting it is sufficiently realistic to study \Lya emission from the cosmic web.}
\label{fig:laeprops}
\end{figure}

In Figure~\ref{fig:laeprops}, we show four observable properties and relations for \Lya emitting galaxies. None were used for calibration, enabling us to use them to assess the performance and realism of the emission and rescaling model. In the first panel, we show the UV luminosity function with (orange) and without (blue) accounting for dust, as treated in Equation~\ref{eq:uvdust}. The dust-free UV luminosity function has a nearly constant slope down to magnitudes of M$_{\mathrm{UV}}=-22$ after which a substantial decline sets in (not shown). With dust attenuation, this decline sharpens and already sets in at M$_{\mathrm{UV}}=-19$. Red error bars show observational data points from the photometric redshift sample in~\citet{Mehta17} at $z=2$. Generally, the simulated UV luminosity function is in reasonable agreement with the observational data, except for the high luminosity end beyond M$_{\mathrm UV}=-20$ where the simulated LF drops off faster than observed.

The second panel shows \Lya luminosity as a function of the UV magnitude. The median (orange line) indicates a positive correlation between UV and \Lya luminosity, that begins to flatten towards higher UV luminosities. This is a consequence of our dust model and empirical calibration, which increasingly suppress \Lya emission from massive, dust rich galaxies. 

This suppression can be seen clearly in the third panel, showing the \Lya escape fraction as a function of stellar mass. Here we include only galaxies with L$_{\mathrm{Ly}\alpha}> 3\cdot 10^{41}$\ergs, to be roughly consistent with the data selection function. We calculate the escape fraction as the ratio between the scattered \Lya luminosity after rescaling and the intrinsic \Lya luminosity before rescaling ignoring contributions from excitations, which is closest to the methodology used to infer the \Lya escape fraction in observations. The \Lya escape fraction is high for low-mass galaxies, approaching unity below $10^{8}$\msun, before rapidly dropping to only ${\sim}10$\% by $10^{9}$\msun, and reaching ${\sim}2$\% for $10^{10}$\msun. In red, we show observations for the escape fraction from~\cite{Weiss21} based on emission-line galaxies at $z\sim 2$~\citep{Bowman19}, broadly consistent with other observations~\citep{Hayes10, Ciardullo14, Sobral17, Snapp-Kolas22}.

In~\citet{Weiss21} the escape fraction is estimated using the flux ratio of \Lya to dust-corrected H$\beta$, adopting $f_{\mathrm{esc}}^{\mathrm{Ly}\alpha}=1/23 \frac{\Lya}{\mathrm{H}\beta}$. This relation is derived assuming all emission arises from recombinations, and neglecting minor changes due to the temperature dependence of the ratio between the \Lya and H$\alpha$ recombination coefficients. The reasonable agreement in comparison to observations suggests that our model primarily captures dust attenuation (as intended) rather than, e.g., modifying effective galaxy star formation rates. When including emission from excitations, we obtain a slightly steeper relation, which is still broadly consistent with observations.

The last panel shows the distribution of rest-frame equivalent widths, for all galaxies with L$_{\mathrm{Ly}\alpha}> 3 \cdot 10^{41}$\ergs, to best match the selection in the observational data. The distribution is positively skewed and unimodal, peaking around $25$\angstrom. $68\%$ of emitters have REW $\ge 20$\angstrom while the median value is $36$\angstrom. The tail follows an exponential decay similar to the form found in observations by~\cite{Oyarzun17}.

In all four panels we account for only the most basic and zeroth order selection effects. These comparisons will in the future benefit from more sophisticated mocks. Nevertheless, our overall result is that the calibrated \Lya emission model is reasonably realistic and broadly consistent with available $z=2$ data.

\subsection{The physical origin of Lyman-alpha emission}
\label{sec:mechsandorigins}

\begin{figure*}
\centering
\includegraphics[width=0.98\textwidth]{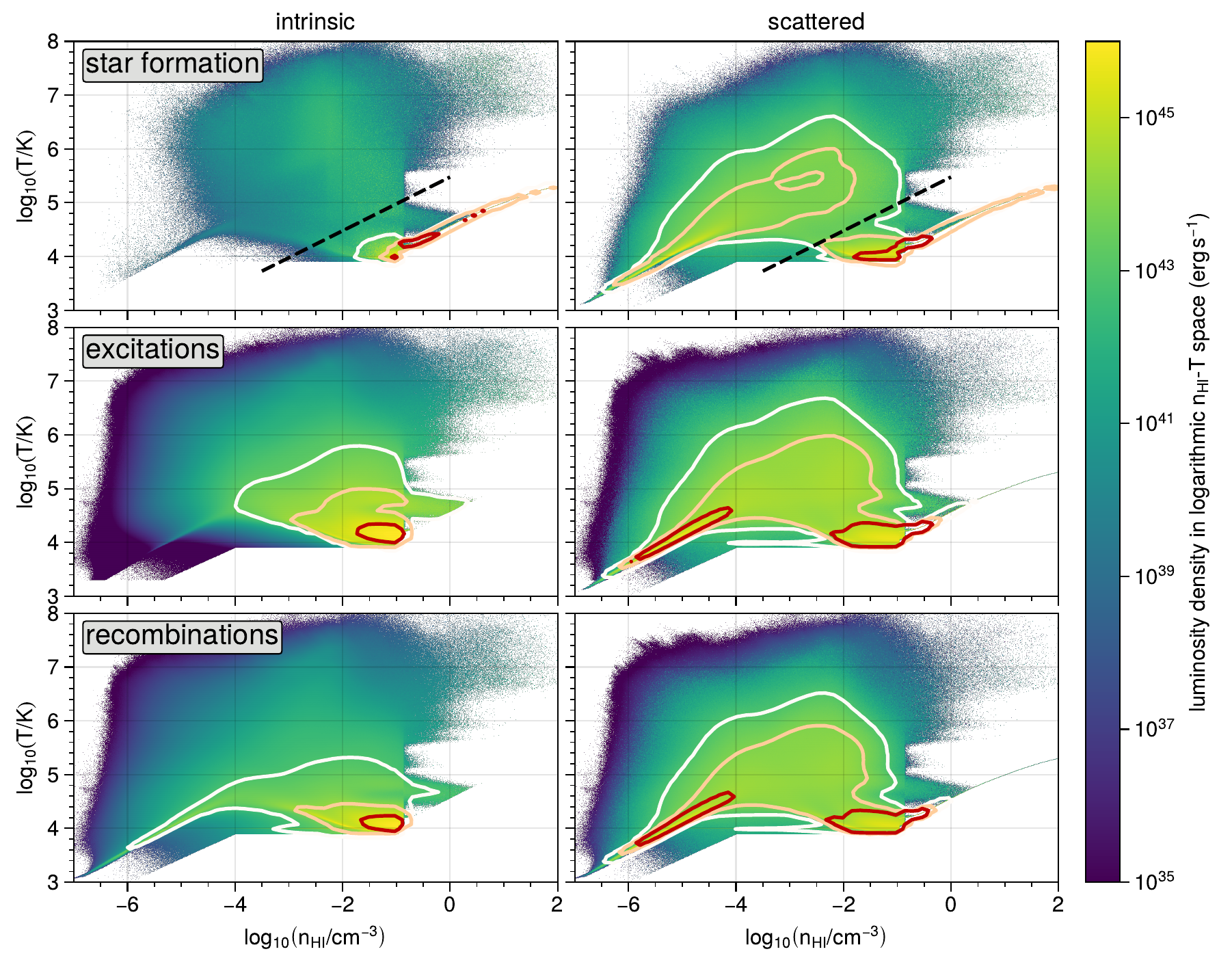}
  \caption{Total \Lya emission as a function of gas density and temperature at $z=2$. Different rows represent the different emission mechanisms: nebular emission sourced by star formation (top), collisional excitation (middle), and recombination (bottom). The left column adopts the physical gas state at the time of emission (``intrinsic''), whereas the right column takes the local gas properties at the time of last scattering (``scattered''). The color map represents the luminosity density in logarithmic $\mathrm{n}_\mathrm{HI}$-$\mathrm{T}$ space. The contours show the brightest phase space regions containing 99, 95 and 50 percent of the total luminosity in red, orange, and white. In the case of nebular emission sourced by star formation, the bulk of emission comes from actively star-forming gas. A small contribution comes from other environments that stellar populations migrate into. For emission from excitations, dense gas with n$_{H}>10^{-2}$\,cm$^{-3}$ at $T\sim 10^{4.2}$\,K dominates the overall budget. A similar behavior is found for recombinations, albeit at lower temperatures. After radiative transfer, most photons illuminate cold gas at lower densities when compared to their point of emission.}
\label{fig:phasespace_allmechs}
\end{figure*}

In Figure~\ref{fig:phasespace_allmechs} we show density-temperature phase space diagrams, weighted by the total \Lya emissivity. Different rows represent the various emission mechanisms: nebular emission from stellar populations (top), collisional excitation (middle), and recombination (bottom). In the left panels, we show the emissivities for the intrinsic photons, i.e., the gas state at the site of emission. In the right panels, the gas state at last scattering is instead shown, i.e.\ the gas state at the location where the \Lya photon last interacts before escaping towards the observer. The contours show the brightest regions of phase space containing 99, 95 and 50 percent of the total luminosity.

The first row shows the emissivities from nebular emission sourced by stellar populations. As expected, most of the emission originates in the star-forming gas above $n_{\mathrm{H}}\geq 0.13$\,cm$^{-3}$. However, nearly the entire phase space contains stellar emission due to a small fraction of stars that have migrated out of star-forming regions (left panel). At last scattering (right panel), a significant fraction of photons have been redistributed out of star-forming regions. We quantify this by integrating the luminosity in the cold-dense region of the phase diagram, defined as the area below the black dashed line in Figure~\ref{fig:phasespace_allmechs}. We find that $99.3$\% of the intrinsic emission stems from this region, dropping to $69.9$\% after radiative transfer, indicating the large spatial redistribution of photons originating around stellar populations.

For diffuse emission from excitations and recombinations, emission primarily stems from a similar region of the phase diagram: dense, cold gas. Emission for recombinations is also significant from low-density regions characteristic of the intergalactic medium, where ionization is powered by the UVB radiation field. On the other hand, emission from collisional excitations quickly drops faster towards low densities. Comparing the two columns, we see a significant redistribution of photons in phase space when comparing intrinsic and scattered photons. After accounting for radiative transfer, photons more homogeneously sample larger regions of phase space. A significant fraction of the total \Lya emission for both recombinations and collisional excitation arises, after scattering, from the cool, low-density IGM.

\begin{figure*}
\centering
\includegraphics[width=1.0\textwidth]{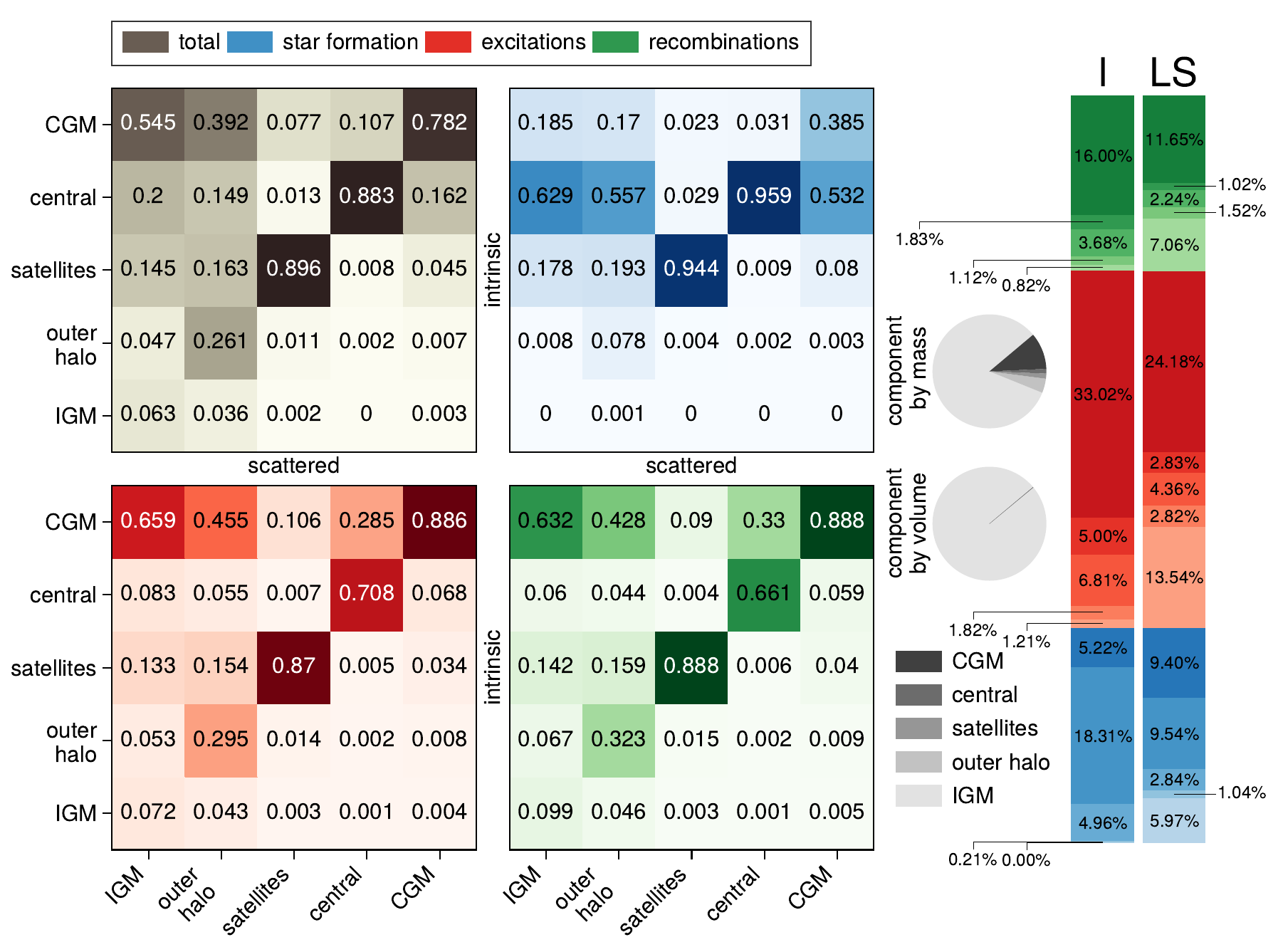}
  \caption{Global \Lya luminosity budget at $z=2$, decomposed by emission mechanism and spatial component. In the bar charts on the right, we show the contributions due to emission from intrinsic photons (``I'') and from scattered photons at last scattering (``LS''). For comparison, the pie charts show the contribution of each spatial component to the global mass and volume budget. On the left, each matrix shows the redistribution of emission arising in a given spatial component by radiative transfer. Here each emission mechanism is considered separately: the total (gray), nebular emission around stellar populations (blue), excitations (red), and recombinations (green). Within each column, entries add up to unity. This way, each number in a column represents the fraction of scattered photons originally emitted by some component given by row.}
\label{fig:mechs_components_overview}
\end{figure*}

In Figure~\ref{fig:mechs_components_overview} we give an overview of the global \Lya luminosity budget at $z=2$, considering the different emission mechanisms and how photons are redistributed by scattering between spatial components. In the bar charts on the right, we show the relative importance of emission mechanisms and spatial components reaching the observer. The relative fractions of intrinsic emission (``I'') stemming from stars, excitations, and recombinations are shown in blue, red, and green respectively, with corresponding total contributions of $29\%$, $48\%$ and $23\%$, making collisions the most important emission mechanism for the overall volume. Note that if we exclusively consider the denser environments of filaments, with dark matter overdensities from $3$ to $30$, emission from stars starts to dominate the luminosity budget (not shown). Each colored area is subdivided into five regions, corresponding to the five distinct spatial components: CGM, central, satellites, outer halo, and IGM, with increasing transparency. In addition, the two gray pie charts show the global fraction of each spatial component, regardless of emission mechanism, by mass and volume.

Focusing on observable \Lya photons (at last scattering; ``LS''), while the IGM encompasses $99.8\%$ of the entire volume of the  Universe and contains $83\%$ of all matter, only ${\sim}16\%$ of emission originates here. On the other hand, the CGM is the single largest spatial component for all three emission mechanisms, contributing more than half of all emission, with only $10$\% of matter in the Universe. Central galaxies contribute a similar fraction of emission due to stars, but add little via recombinations and collisions, with a combined ${\sim}14\%$ of all emission. The outskirts of halos and satellites combined contribute ${\sim}18\%$ to the overall emission, signifying sub-dominant but non-negligible spatial components.

On the left of Figure~\ref{fig:mechs_components_overview}, we show four matrices for the emission mechanisms: all combined (i.e.\ the total, in gray), nebular emission due to star formation (blue), excitation (red), and recombination (green). The matrices reveal the degree to which photons emitted in a certain component are redistributed to another due to radiative transfer effects. Each column is normalized to unity. As a result, the numbers give the fraction of \Lya luminosity emerging (i.e.\ last scattering) from a given spatial component, which originated (i.e.\ intrinsically) elsewhere. For satellites, centrals, and the CGM, most of the observed emission does originate from those components themselves. However, this does not hold for the outer halo and the IGM, where most of the emission comes from the circumgalactic medium of galaxies (for recombinations and excitations) and central galaxies themselves (for stellar populations). Specifically, $94$\% of all \Lya radiation reaching us from the IGM -- that is, the large-scale cosmic web -- actually does not originate there, stressing the importance of the \Lya radiative transfer modeling in the IGM.

\begin{figure*}
\centering
\includegraphics[width=1.0\textwidth]{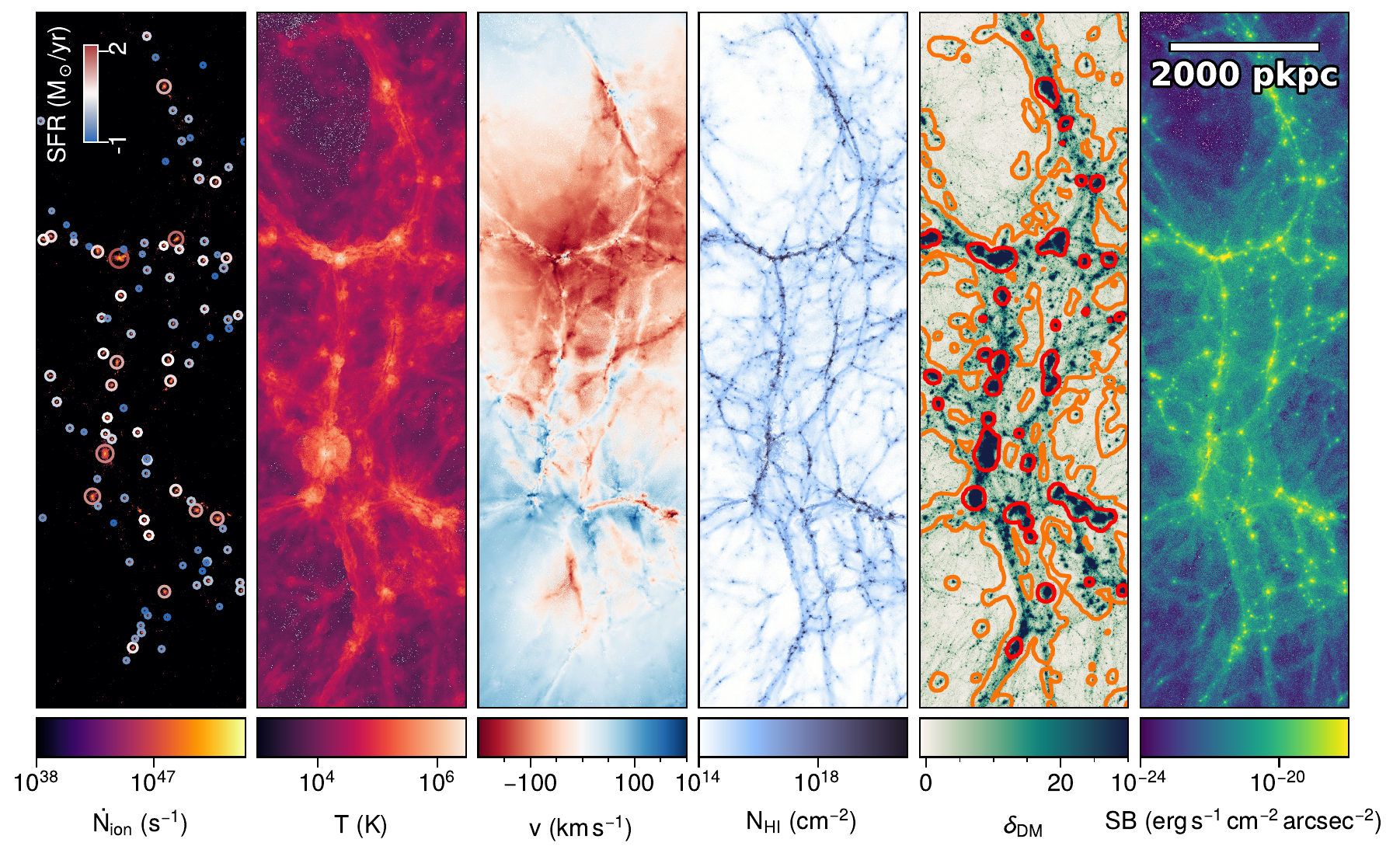}
  \caption{Physical properties shaping the \Lya emission and scattering for the large zoom-in region shown in Figure~\ref{fig:overviewplot}, a subset of TNG50 at $z=2$, measuring ${\sim}3 \times 10$\,pMpc across and projecting $2.3$\,pMpc along the line-of-sight. The first panel shows the integrated ionizing photon rate from stellar contributions. Circles indicate the virial radii of objects with a star-formation rate above $0.1$\msun\,yr$^{-1}$. The color of the circles reflects the amount of star-formation. The second and third panels show the mass-weighted temperature and line-of-sight velocity, respectively. The fourth panel shows the neutral hydrogen column density. The fifth panel shows the dark matter overdensity with a Gaussian smoothing of $\sigma=100$\,pkpc. We also included contours of 3 (30) in orange (red) projecting the dark matter overdensity maximum across the projection. In the last panel, we show the resulting \Lya surface brightness map, given our fiducial model and a full treatment of the \Lya resonant scattering, highlighting how \Lya emission traces the cosmic web.}
\label{fig:slice_physicalproperties}
\end{figure*}

The total luminosity density of \Lya, based on our fiducial emission and radiative transfer models applied to TNG50 at $z=2$, is $\dot{\rho}_{\mathrm{Ly}\alpha}=4.4\cdot 10^{40}$\ergs\,cMpc$^{-3}$. Central galaxies emit a quarter of the total luminosity density with $1.1\cdot 10^{40}$\ergs\,cMpc$^{-3}$. Moreover, $98$\% of all emission originates within halos, with only $9.1\cdot 10^{38}$\ergs\,cMpc$^{-3}$ originating in the IGM\@. When considering the luminosity density at last scattering, this picture changes significantly: central galaxies retain only half of their emission upon reaching the observer with $5.8\cdot 10^{39}$\ergs\,cMpc$^{-3}$ and halos only contain $73$\% of the luminosity budget scattering there last. The remainder is redistributed into the IGM, boosting its luminosity density by an order of magnitude to $1.2\cdot 10^{40}$\ergs\,cMpc$^{-3}$.

\subsection{Visual inspection of the large-scale topology, origin, and physics of the Lyman-alpha cosmic web}
\label{sec:visualinspection}

Figure~\ref{fig:slice_physicalproperties} visualizes a large $3 \times 10$\,pMpc region selected to include both large-scale and small-scale filamentary gas structures (white rectangle in Figure~\ref{fig:overviewplot}). From left to right we show: the ionizing photon rate, temperature, line-of-sight velocity, neutral hydrogen column density for a slice depth of $5.7$\angstrom, the dark matter density distribution, and the resulting \Lya surface brightness map, based on our fiducial model and including radiative transfer effects.

The ionizing photon rate projection is sparsely populated, showing that the stars within galaxies trace only the highest overdensities. These sites of ionizing photon production are correlated with the filaments visible in neutral hydrogen column density, for example. In the temperature, line-of-sight velocity and neutral hydrogen density panels, we can clearly recognize filamentary structures, across a variety of length scales, from shorter $\sim 100$\,kpc filaments to larger $\sim$\,Mpc long filaments. In general, filaments have a velocity relative to their background, are relatively cold and optically thick. They are often surrounded by hot regions powered by feedback and shocks. The dark matter overdensity field traces the neutral hydrogen column density. However, while the dark matter has a more pronounced and clumpy substructure, it does not show the distinct string-like filaments visible for neutral hydrogen. Most importantly, the \Lya surface brightness clearly traces the filamentary structure, suggesting that the cosmic web in \Lya emission is closely linked to the underlying gas and dark matter distributions on large scales. However, \Lya does not have a simple one-to-one mapping with gas density, as visible for instance in the more diffuse and less pronounced edges and boundaries in comparison to the neutral hydrogen distribution.

\begin{figure}
\centering
\includegraphics[width=0.47\textwidth]{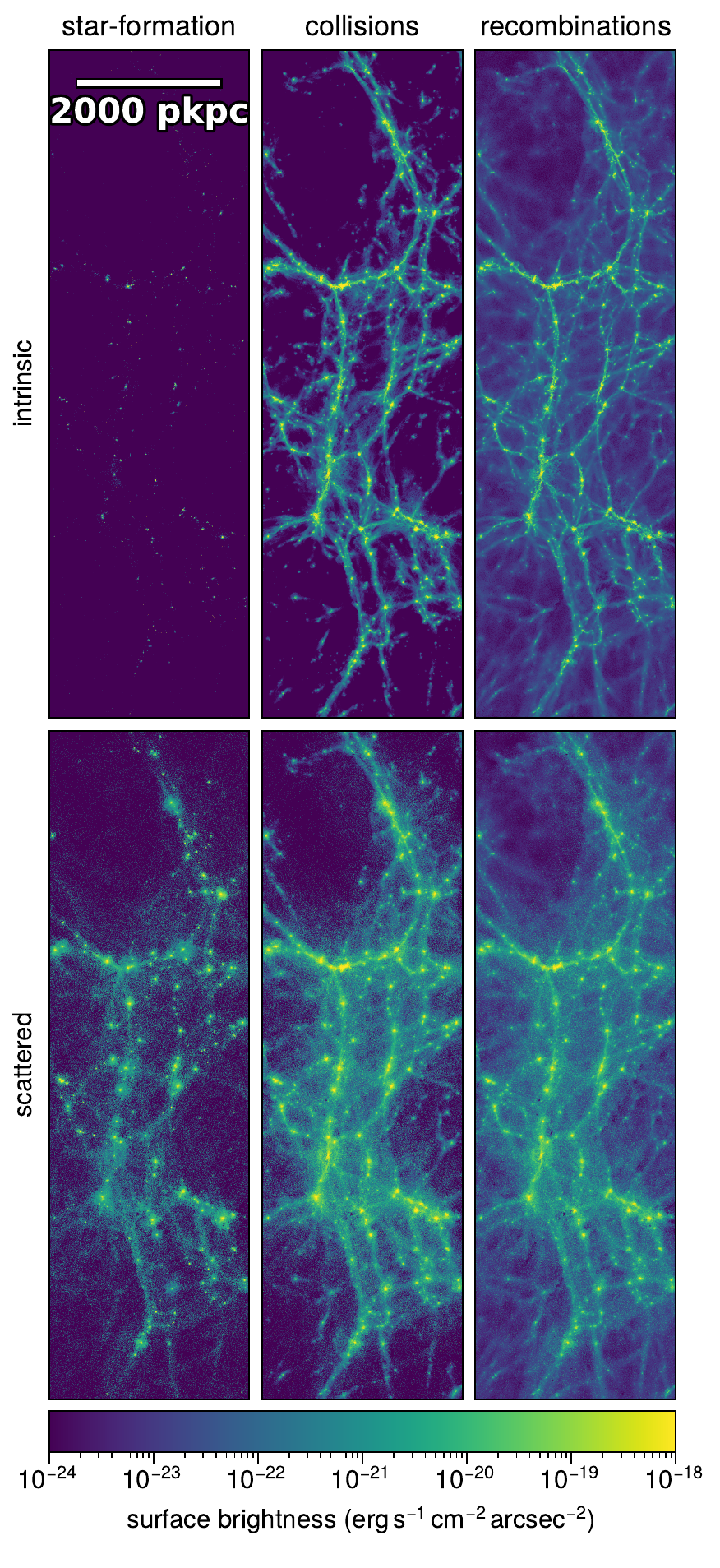}
  \caption{Decomposition of the \Lya surface brightness on large scales, splitting into the three different emission mechanisms. We show the same region of space as before, which has numerous visible cosmic web filaments. The upper row shows the intrinsic emission, i.e.\ \Lya photons contributions at the location they are originally produced. In contrast, the lower row shows the observable view of \Lya photons after they have undergone our full radiative transfer treatment for scattering. For intrinsic emission, the cosmic web is traced by collisions and recombinations, while emission from stars only traces spatially compact areas representing galaxies. Emission from collisions is concentrated towards higher densities compared to recombinations, where surface brightness drops more gradually towards regions of low densities. For the observable surface brightness maps, photons from filaments originate in large part from nearby gas. In the case of emission from stars, significant emission is redistributed from galaxies into filaments.}
\label{fig:slice_mechs}
\end{figure}

In Figure~\ref{fig:slice_mechs}, we study the \Lya surface brightness (SB) projection for this same region of space in more detail. We decompose the total emission into the contributions from the three emission mechanisms (different columns). The top panels show intrinsic emission, i.e.\ \Lya photons where they are initially emitted. The top left panel shows nebular emission sourced by star-formation, where we see sparsely distributed emission stemming from massive star-forming galaxies, similar to the projection of ionizing photon rates in Figure~\ref{fig:slice_physicalproperties}. There is no strict proportionality of ionizing photon rates and \Lya emission (see Equation~\ref{eq:LyaSF}) due to our application of the empirically calibrated rescaling model. Collisions and recombinations (top  middle and top right panels) both roughly follow the distribution of neutral hydrogen and dark matter. For collisions, we see a higher contrast with strong emission in the filamentary structure that quickly drops below $10^{-24}$\sbunits, compared to recombinations that maintain a higher surface brightness, even in voids.

The bottom row of panels in Figure~\ref{fig:slice_mechs} shows the surface brightness maps after accounting for the radiative transfer, i.e.\ after processing these photons to account for the resonant scattering process. The most striking difference arises in star-formation sourced emission. While intrinsically confined to within galaxies, after scattering these same photons illuminate the filaments of the cosmic web. While photon redistribution for collisions and recombinations is less pronounced, the spatial diffusion of photons increases the surface brightness of filament outskirts. In terms of peak surface brightness and contrast, filaments are most pronounced due to \Lya emission from collisional excitation. In this case, radiative transfer effects lead to somewhat fuzzier filaments due to spatial diffusion.

All mechanisms independently give rise to surface brightnesses above $10^{-19}$\sbunits around massive objects, but filamentary structures above \mbox{$3\cdot 10^{-20}$\sbunits} are only visible in the proximity of massive galaxies and cosmic web nodes. In this case, AGN provide additional photoheating and -ionization. With decreasing surface brightness, larger connected filamentary structures appear. Thin filaments extending more than a physical Megaparsec are easily identifiable by eye at ${\sim} 10^{-22}$\sbunits due to collisional excitations and recombinations. We expand on the quantitative brightness of \Lya filaments, and their observational detectability in the following sections.

\begin{figure*}
\centering
\includegraphics[width=1.0\textwidth]{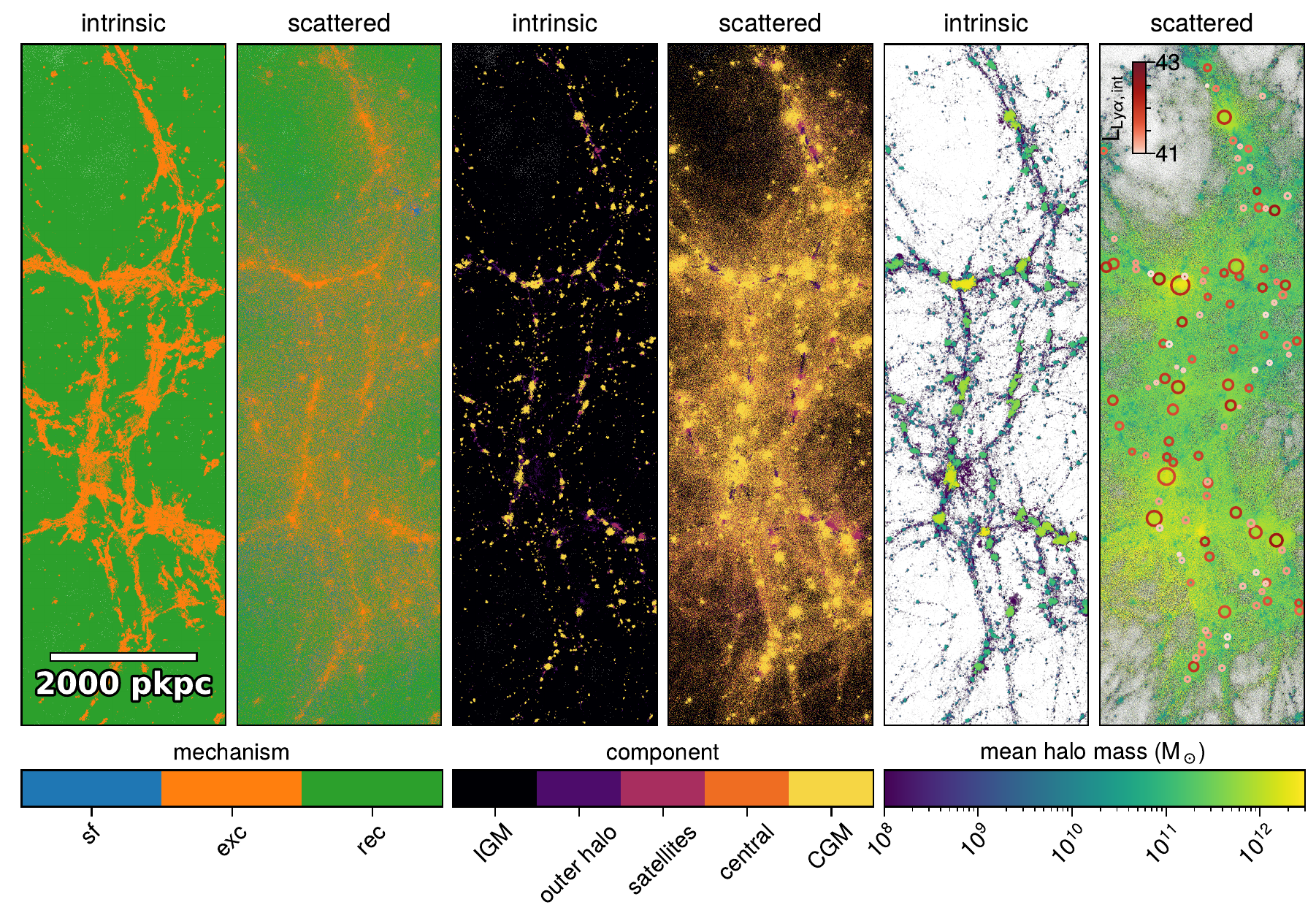}
  \caption{Visualization of the dominant \Lya emission mechanism (left two panels), spatial component (middle two panels) and mean contributing dark matter halo mass (right two panels). The first two pairs show the mechanism/component dominating each pixel, based on \Lya luminosity. For the last pair, the luminosity-weighted halo mass is shown. In the left panel of each pair, we show the dominant mechanism/component and mean halo mass for intrinsic photons, i.e.\ at the location of their original emission. In the right panel of each pair, we show the dominant property of a given pixel after including radiative transfer effects, i.e.\ photons are included at their observable (last scattering) locations. The three emission mechanisms (star-formation in blue, excitation in orange, recombination in green) and five spatial origins (IGM in black, outer halo in purple, satellites in magenta, central galaxies in orange, and CGM in yellow) are the same as previously. Note that for the last two panels, photons which originate outside of dark matter halos -- a sub-dominant component -- are not considered. For the last panel, we encircle galaxies with intrinsic Lyman-alpha luminosities above $10^{41}$\ergs, with color representing the intrinsic luminosity, and radius equal to the virial radii. Our key finding is that the largest \Lya filaments are illuminated by photons which are: predominantly created via excitations (orange, left panels) within the circumgalactic media (yellow, middle panels) of intermediate mass halos (green, right panels).}
\label{fig:zoomin_rt}
\end{figure*}

In Figure~\ref{fig:zoomin_rt}, we visually decompose the total \Lya surface brightness of the same region of space into the dominant emission mechanism (left panels), spatial component, i.e.\ origin (middle panels), and contributing mean halo mass (right panels). In all cases, pixels of the images are colored by the mechanism/component/halo mass which contributes most of the \Lya luminosity to that pixel. The left panel of each pair shows the intrinsic emission, while the right panel of each pair instead shows the result after scattering, i.e.\ the observable \Lya emission after accounting for the radiative transfer process. With respect to the emission mechanism in the intrinsic case (left-most panel), we find excitations within filaments and recombinations outside of them to dominate, and the separation is clear. There exist only small regions within filaments situated within the most massive halos where star-formation dominates. After radiative transfer, most filaments remain dominated by excitations, and the regions dominated by star-formation sourced emission enlarge. Most importantly, emission in filament outskirts and in voids is now dominated by either recombinations or stellar emission rather than recombinations.

The middle panels of Figure~\ref{fig:zoomin_rt} show the dominant component, i.e.\ spatial origin, before and after radiative transfer. Before radiative transfer, we see that most of the area around the filaments is dominated by the CGM of galaxies, but there remain large unbound areas that are dominated by the IGM. Occasionally, regions within filaments occur where satellites, centrals or outer halo origins dominate. After radiative transfer, most of the surface brightness of cosmic web filaments is dominated by emission originating from the circumgalactic medium of galaxies, residing both within and outside of filaments. This is a key result of our study. Only small regions remain where emission originating from the IGM dominates.

The fifth and sixth panels of Figure~\ref{fig:zoomin_rt} show the mean dark matter halo mass, from which photons dominate, weighted by \Lya luminosity. Photons which originate outside of all halos are not considered in this case. In general, significant fractions of the filaments are filled by gas from massive halos ${\gtrsim}10^{10}$\msun, and the intrinsic emission from these relatively high-mass halos is directly responsible for emission from the filament regions. The role of emission from halos as a function of mass changes after we account for radiative transfer effects (right-most panel). In this case, we find that the emission from massive galaxies often overshadows that of the smaller halos. This is facilitated by a significant photon flux of massive halos reaching and scattering off the CGM and IGM surrounding smaller halos and scattering into, and then out from, smaller halos often hundreds to thousands of physical kiloparsecs away \citep{Byrohl21}.

\subsection{Surface brightness distributions}
\label{sec:sbdist}

\begin{figure}
\centering
\includegraphics[width=0.49\textwidth]{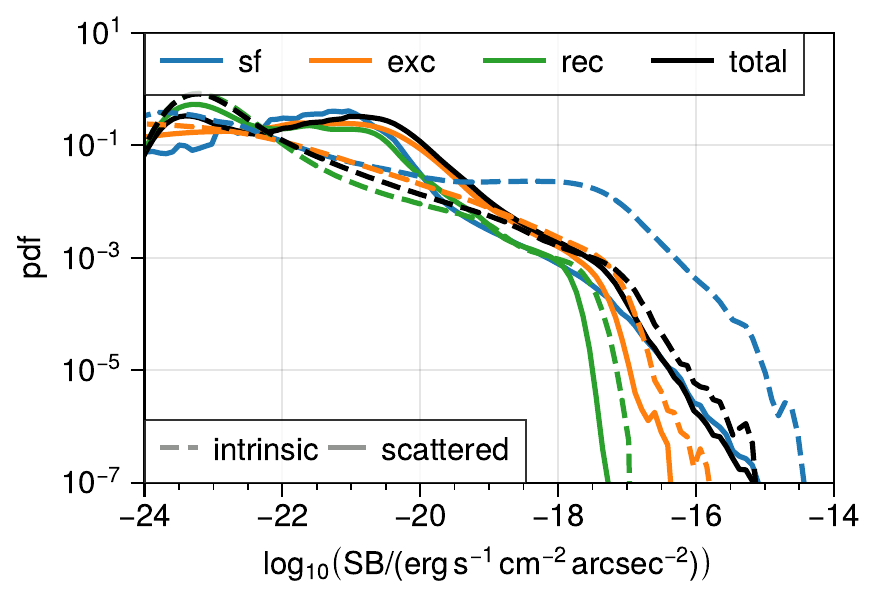}
  \caption{Distribution of \Lya surface brightness values at $z=2$. The probability density is given per logarithmic bin of surface brightness (SB). We decompose the total distribution (black) into its three emission mechanisms: nebular emission sourced by star formation (blue), collisional excitation (orange), and recombination (green). In all cases, the intrinsic case neglecting scattering is indicated by dashed lines, while the observable SB values after accounting for radiative transfer effects are shown by solid lines. The overall distribution is multimodal, indicating a number of contributing components. The absolute SB values $>-17$\sbunits arise due to stars, while intermediate SB values between $10^{-20}$ and $10^{-17}$\sbunits show a complex behavior, but are dominated by excitation. At low SB values (between $10^{-22}$ and $10^{-20}$\sbunits) a mixture of all three mechanisms shapes the overall distribution.}
\label{fig:sbpdf_mechs}
\end{figure}

We now quantify the sky area covered by \Lya emission at different surface brightness levels. In Figure~\ref{fig:sbpdf_mechs}, we show the fraction of pixels with a certain surface brightness for the respective emission mechanism before (i.e.\ intrinsic; dotted lines) and after (i.e.\ scattered; solid lines) radiative transfer. For collisions and recombinations (orange and green) the intrinsic surface brightness distribution drops sharply at around $10^{-17.5}$\sbunits with collisions extending to SB values higher by roughly a factor of 3. If we further decomposing the distribution based on the existence of an AGN radiation field at a given location, we find the high surface brightness tail for collisions and recombinations is dominated by emission from gas experiencing additional photoheating and -ionization (not shown). At lower surface brightness, recombinations and collisions show a similar qualitative trend. For star-formation sourced emission (blue), a substantial number of pixels in the intrinsic surface brightness distribution extend to significantly higher values compared to collisions and recombinations.

Comparing dashed (for intrinsic photons) and solid (for scattered photons) lines, we find the largest impact of radiative transfer at the high luminosity end where the occurrence of peak SB values decreases by a factor of a few, and at surface brightness values around $10^{-21}$\sbunits where it is enhanced by an order of magnitude when radiative transfer is applied. That is, the importance of \Lya resonant scattering is actually maximal at the SB values which correspond to the cosmic web filaments, making a full radiative transfer treatment essential. Regarding the three emission mechanisms, the impact of radiative transfer is largest for emission from stellar populations. Before scattering, the distribution is relatively flat, but after applying radiative transfer large fractions of the previously \Lya dim sky are boosted in brightness, such that the surface brightness distribution roughly resembles the distribution of the other two mechanisms.

\begin{figure}
\centering
\includegraphics[width=0.49\textwidth]{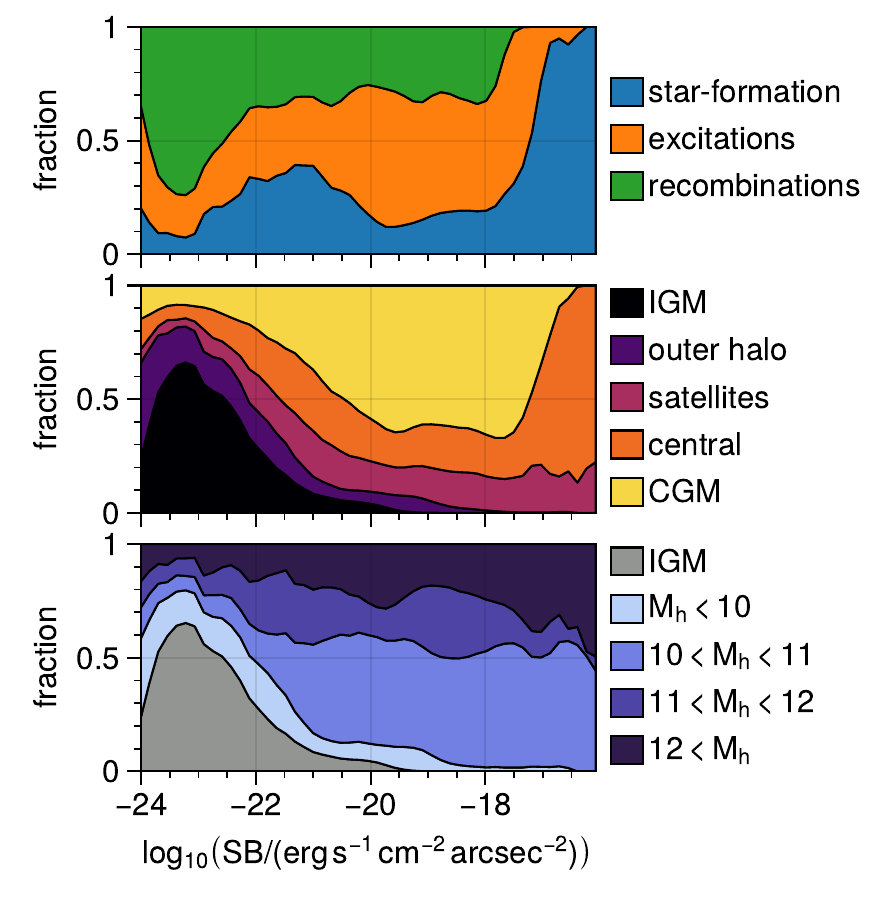}
  \caption{The fraction of \Lya luminosity, as a function of surface brightness, split by emission mechanism (top), component i.e.\ spatial origin (middle), and originating halo mass (bottom). We show the fractions stacked vertically, such that they cumulatively sum to unity at each SB value, and the relative colored area represents the relative fraction of that mechanism/component/halo mass. M$_{\mathrm{h}}$ is $\log_{10}$ of the total halo mass in a sphere of a density $200$ times the critical density, in units of M$_{\odot}$. The largest SB values are dominated by emission due to stellar populations in central galaxies with halo masses between $10^{10}$ and $10^{11}$\msun. At lower SB values, collisional excitations start to quickly dominate, while emission from the CGM in halos above $10^{10}$\msun remains the dominant contributor above $10^{-21}$\sbunits. Only below this level does emission from recombinations in the IGM eventually dominate.}
\label{fig:sbpdf_fracs_all}
\end{figure}

In Figure~\ref{fig:sbpdf_fracs_all} we quantify the relative luminosity contributions for different emission mechanisms (top), components (i.e.\ spatial origin, middle), and contributing dark matter halo mass ranges (bottom) as a function of surface brightness. The colored area at a given surface brightness reflects the relative fraction of each legend item, stacked vertically. The results are shown for the scattered photons, i.e.\ after accounting for radiative transfer effects.

The upper panel shows the contribution by emission mechanism. We find star-formation (blue) sourced \Lya radiation dominates SB values above $10^{-18}$\sbunits. For smaller values, star-formation remains a relevant emission channel, contributing $10$\% to $35$\%. Between $10^{-21}$ and $10^{-18}$\sbunits, collisional excitation (orange) sources most of the observed surface brightness. For values below $10^{-21}$\sbunits, recombination (blue) starts dominating.

The middle panel shows the fraction of \Lya luminosity, as a function of surface brightness, originating from each spatial component. Luminosity originating from central galaxies (orange) dominates at high SB values, and down to $10^{-18}$\sbunits, coinciding with the trend of star-formation as in the upper panel. Between $10^{-20}$ and $10^{-18}$\sbunits, luminosity from the CGM (yellow) contributes more than $60$\%. Below $10^{-20}$\sbunits, the contribution from the CGM declines, and IGM (black) contributions gradually grow until the IGM eventually becomes the most important origin below $10^{-22}$\sbunits. Satellites and outer halos (magenta and purple) contribute ${\sim}20-30\%$ to the budget throughout the full dynamic range of surface brightness.

The lower panel shows the luminosity fraction contributed by emission from halos as a function of halo mass. After we account for radiative transfer effects and consider \Lya photons as they would be actually observable, we find that intermediate-mass halos are crucial \Lya sources. Specifically, between $10^{10}$ and $10^{11}$\msun they contribute more than $50\%$ to \Lya filament luminosity, down to $10^{-21}$\sbunits.

We can evaluate these same contributions for intrinsic photons, i.e.\ neglecting radiative transfer (not explicitly shown). If we do so, this same halo mass range dominates at high SB above $10^{-19}$\sbunits, while the other three halo mass bins have roughly equal contributions below, and the IGM becomes the major contributor at the lowest surface brightness levels, below $10^{-21}$\sbunits. Similarly, stellar contributions quickly approach zero, reaching $<10$\% below $10^{-19}$\sbunits, while the CGM contributes significantly less to the luminosity budget, and emission from the IGM dominates below $10^{-20}$.

Finally, when including radiative transfer, we can also consider the spatial component at the point of last scattering (also not explicitly shown), rather than at the origin. When doing so, we find that there is a sharp transition at ${\sim}10^{-19.5}$\sbunits below which nearly all photons are scattered by the IGM\@.

That is, we find that the \Lya cosmic web in emission is illuminated predominantly by photons which (i) originate within central galaxies and the CGM of intermediate-mass halos, and (ii) scatter into the IGM before reaching the observer.

\begin{figure}
\centering
\includegraphics[width=0.49\textwidth]{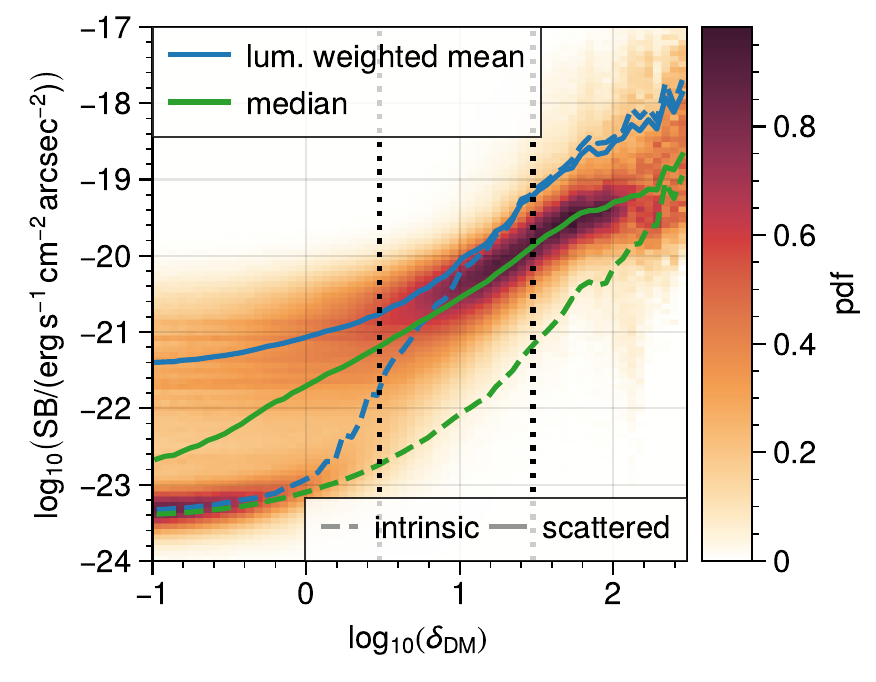}
  \caption{Distribution of \Lya surface brightness as function of dark matter overdensity field, where we smooth the with a Gaussian $\sigma=100$\,pkpc filter. The distribution is normalized to unity for each dark matter overdensity. The green line shows the median surface brightness at a given dark matter overdensity, while the blue line indicates the mean. Solid lines show the values for scattered photons, and dashed lines for intrinsic photons. At overdensities characteristic of dark matter filaments outside of halos, $\delta_{\mathrm{DM}}$ from $3$ to $30$ chosen here (see Figure~\ref{fig:slice_physicalproperties}; vertical dotted lines), \Lya SB values have a mean ranging from $10^{-20.7}$ to $10^{-19}$\sbunits and increase rapidly in more dense regions. The median value is commonly a factor of ${\sim}3$ lower. Note that, particularly the median depends on the point spread function of the \Lya surface brightness maps. We currently only impose the initial binning with a resolution of ${\sim}0.5$\,arcsec without further smoothing.}
\label{fig:sbpdf_overdensity}

\end{figure}

In Figure~\ref{fig:sbpdf_overdensity} we show the \Lya surface brightness as a function of dark matter overdensity $\delta_\mathrm{DM}$. We construct the dark matter overdensity field and impose a Gaussian filter with $\sigma=100$\,pkpc. We then project the maximum overdensity in each pixel of the slice and create a two-dimensional histogram together with the \Lya surface brightness for the pixels in the slice. We show the median (luminosity weighted mean) in green (blue), separately for scattered (intrinsic/unscattered) photons as solid (dashed) lines. In the background, we show the probability density of logarithmic surface brightness at given dark matter overdensity.

We find that the observed mean surface brightness is a strictly monotonic function of overdensity starting at ${\sim}10^{-21}$\sbunits for overdensities of a few, characteristic of the cosmic web filaments. This increases to $10^{-18}$\sbunits at overdensities around 200, characteristic of collapsed halos. The mean surface brightness is boosted by around an order of magnitude over the full overdensity range, by redistributing \Lya emission from the densest regions. For scattered photons, the central $68$\% of values around the median show a scatter of roughly $1.5$\,dex across the range of overdensities shown. In comparison, for intrinsic photons this spread grows from $0.5$ to $3$ dex with overdensity. The tight correlation for intrinsic photons at low SB values is set by UVB photoionized hydrogen outside of filaments. This correlation broadens at larger values significantly due to the complex temperature and density structure within halos, due to the lack of strong correlations with the smoothed dark matter field. For scattered photons, the scatter at low surface brightness increases as a fraction of underdense regions are significantly boosted in surface brightness when in proximity to \Lya bright halos.

At high surface brightness the scatter decreases as photon contributions are smoothed out in hydrogen rich, overdense regions. There is a rapid increase between $10^{-23}$ and $10^{-21}$\sbunits where scattered contributions start dominating, over otherwise lower SB values from intrinsic contributions. The surface brightness mean (median) value grows from $10^{-20.8}$ ($10^{-21.2}$) to $10^{-19.2}$ ($10^{-19.9}$)\sbunits from an overdensity of $3$ to $30$.

\subsection{Lyman-alpha filament identification and detectability}
\label{sec:identification}

\begin{figure*}
\centering
\includegraphics[width=1.0\textwidth]{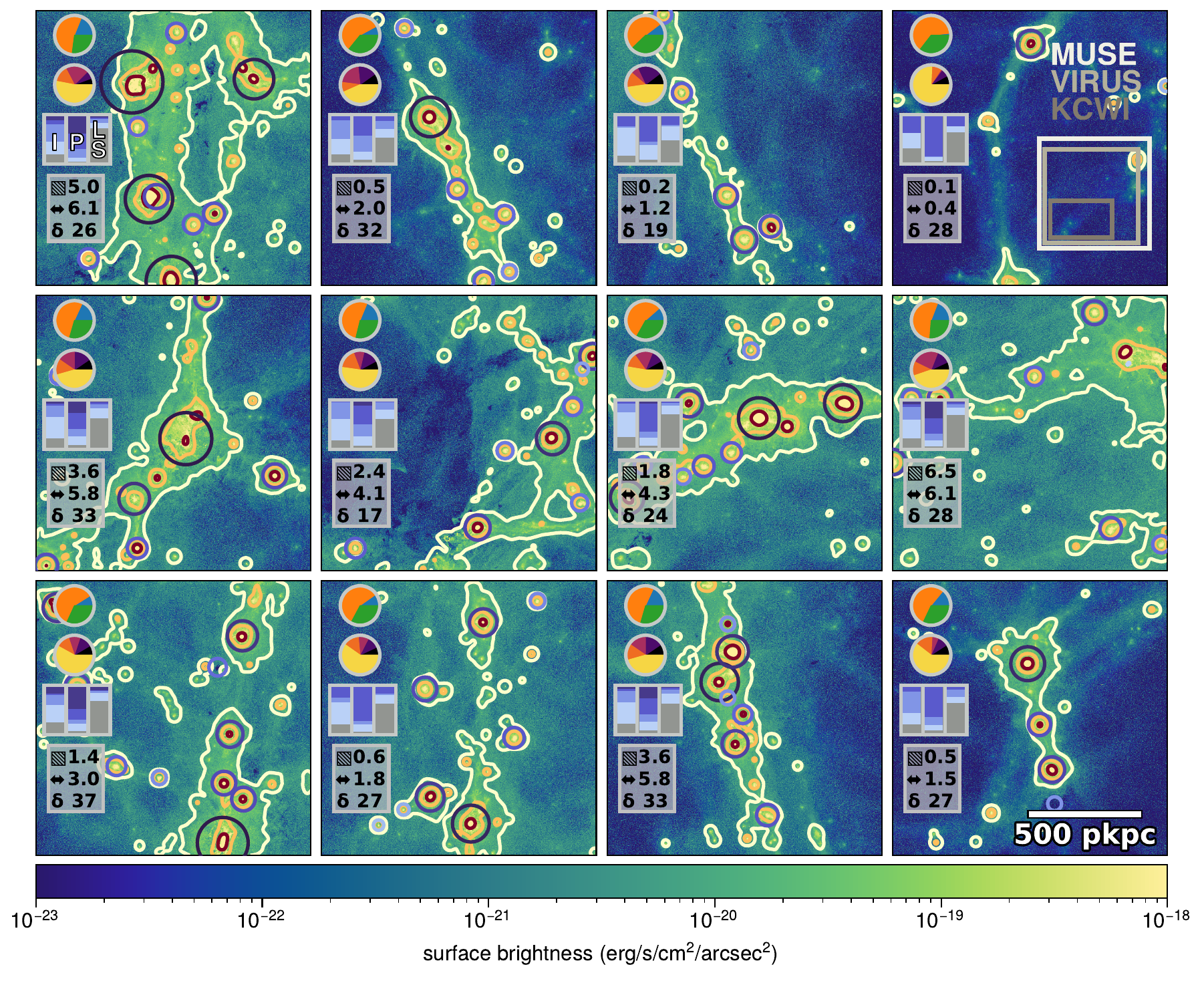}
  \caption{Twelve surface brightness maps containing filamentary structures, all taken from the slice shown in Figure~\ref{fig:overviewplot}. Contours highlight SB values after applying a Gaussian smoothing with a FWHM of $3.5$\,arcsec at levels of $10^{-20}$, $10^{-19}$ and $10^{-18}$\sbunits (light yellow, orange, red). We show circles around all galaxies with \Lya luminosities above $10^{41}$\ergs, color coded by their halo mass and with radius equal to the halo virial radii. In the upper right panel, we show the footprints of integral field units for VLT-MUSE, HET-VIRUS and KECK-KCWI in white, gray and dark gray respectively. Note that for HET-VIRUS only one of 78 installed IFUs is shown. Within each panel, we show properties of the largest visible filament. All properties are evaluated masking out compact regions with SB values above $10^{-19}$\sbunits, thus focusing on the diffuse parts of filaments. We show the relative luminosity contributions for each emission mechanism (first pie chart), component i.e.\ spatial origin (second pie chart) and contributing halo mass (bar chart) with identical color coding to Figure~\ref{fig:sbpdf_fracs_all}. The left bar chart (`I') shows the originating halo mass distribution for intrinsic emission, i.e.\ ignoring any scattering. The middle bar chart (`P') shows the contributions from emission originating for given halo masses. The right bar chart (`LS') also uses the scattered photons, but shows the halo mass at last scattering. Furthermore, the three numbers in each panel show the filament area (in arcminutes$^{2}$), length (in arcminutes) and luminosity-weighted overdensity. The different fields show a rich diversity of filaments in terms of surface brightness, size, morphology, as well as contributing emission mechanism, spatial origin and contributing halo mass. The diffuse regions of filaments ($<10^{-19}$\sbunits) are generally dominated by collisional excitations originating in the CGM, where without radiative transfer low-mass halos ($<10^{10}$\,M$_{\odot}$) would instead dominate.}
\label{fig:zoomin_regions_sbcontours}
\end{figure*}

Next, we aim to identify and characterize \Lya filaments. To do so, we adopt a relatively simple approach. We search for \Lya filaments as connected regions above a certain surface brightness threshold. We label this surface brightness threshold $\mathrm{SB}_{\mathrm{fil},10}$, defined as $\log_{10}$ of the surface brightness in \sbunits. The surface brightness maps are evaluated on a fixed smoothing scale, for which we use a Gaussian filter with a full width half maximum (FWHM) of $\sigma_\mathrm{FWHM, fil}$. Such a smoothing scale would also be imposed on data, in order to identify large-scale structure rather than small-scale details by maximizing available the signal-to-noise ratio. We then measure the filament size $L$ as the maximal distance between any points of a connected region with area $A$. From this region, we also determine the circularity $c= 4A / (\pi L^{2})$ with a value of 0 indicating a one-dimensional line and a value of 1 indicating a perfect circle.

Considering a region as elongated for circularity values below $c<0.5$, we find that \Lya structures with lengths of $L\gtrsim 400$\,pkpc are typically elongated. In the following, we thus commonly adopt a length minimum of $400$\,pkpc for \Lya filaments. For reference, this length threshold roughly corresponds to an area of ${\sim}0.25$\,arcmin for an elongated structure with $c=0.5$ at $z=2.0$. We use a fiducial value of $\sigma_\mathrm{FWHM, fil}=3.5$\,arcsec. We find the number of filaments above $>400$\,pkpc remains nearly constant for smoothing scales $\le 10$\,arcsec, irrespective of the fixed surface brightness threshold considered. Finally, we adopt a fiducial surface brightness threshold of $\mathrm{SB}_{\mathrm{fil}}=10^{-20}$\sbunits as a reasonable value in reach of future surveys. While this results in a robust set of identified filaments, we note that smaller structures with lengths less than $400$\,pkpc are more sensitive to the analysis choices.

In Figure~\ref{fig:zoomin_regions_sbcontours} we show \Lya surface brightness maps of twelve selected zoom-in regions from Figure~\ref{fig:overviewplot}. We select each region by hand, to contain filamentary structures spanning a range of filament sizes and luminosities. We quantify the frequency of such structures below. We include contour levels at $10^{-20}$, $10^{-19}$ and $10^{-18}$\sbunits in light yellow, orange, and red. At $10^{-20}$\sbunits a range of large filamentary structures can be seen, while at higher SB thresholds structures resemble \Lya halos (mostly orange), i.e.\ more circular emission centered on halos, or point sources, as they appear given the smoothing scale (mostly red). On the left side within each panel, we show information for the emission mechanism, spatial component, halo masses (pie charts and bar charts, color coding consistent with other plots) as well as the length, area and dark matter overdensity. These quantities are only evaluated for the diffuse part of filaments with surface brightnesses between $10^{-20}$ and $10^{-19}$\sbunits, and are calculated for the largest filament overlapping the field of view.

While there are variations, particularly for the fraction of \Lya photons sourced by star-formation, we find the collisional excitations and the circumgalactic medium to be the dominant emission mechanisms contributing to the luminosity of filaments. The bar charts, indicate contributing halo masses at emission for intrinsic emission ("I", left), after radiative transfer by origin ("P", center) and last scattering ("LS", right). We find that the majority of \Lya photons originate from halos between $10^{10}$ and $10^{11}$\msun, except in cases where more massive emitters exist in the filament. While most of these photons originate within halos, they scatter off the diffuse IGM before reaching the observer.

\begin{figure}
\centering
\includegraphics[width=0.47\textwidth]{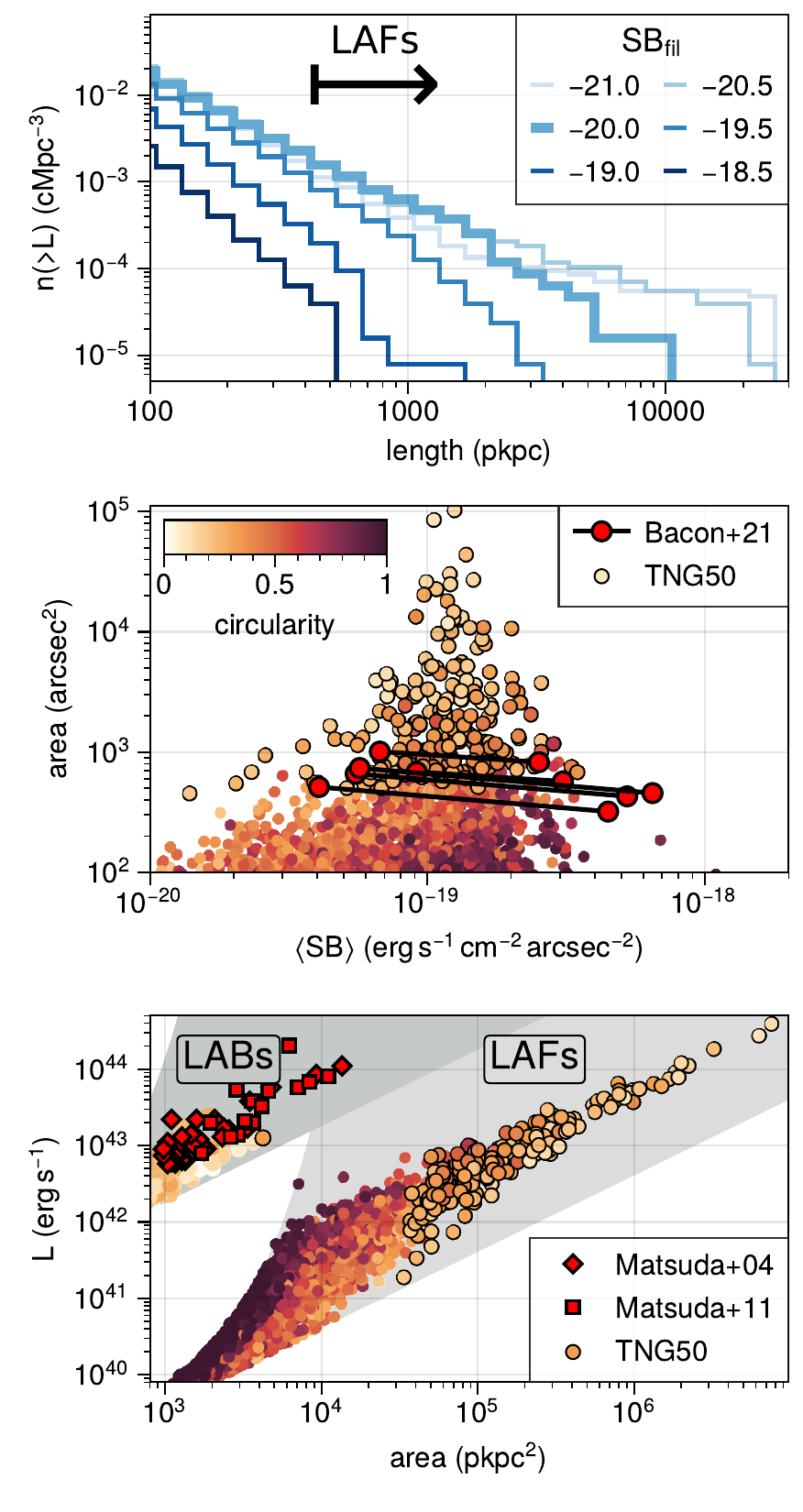}
  \caption{Properties of \Lya filaments (LAFs), which we identify as large, connected areas above a given surface brightness threshold. In the upper panel, we show the cumulative number density of filament lengths at varying surface brightness thresholds in \sbunits (the thickest blue line indicates our fiducial choice of SB$_{\mathrm{fil}}=-20$). The middle panel shows a scatter plot of filament area (y-axis) versus average surface brightness (x-axis). Color shows the circularity parameter, with lighter colors indicating more elongated, i.e.\ filament-like structures. Despite important differences in the detection method, we make a face-value comparison of the observed filaments from \citet{Bacon21}, shown with red circles. Each observational detection is plotted as a pair of two markers: (i) at the redshift of detection (left symbol), and: (ii) with a rescaled surface brightness and area under the assumption that the same filament, with constant luminosity and physical size, was present at $z=2$ (right symbol). The lower panel shows the relationship between \Lya object luminosity and projected area. Shaded regions show allowed regions, for two different configurations of smoothing FWHM and surface brightness threshold (see text). The upper gray area shows LABs ($1.4$\,arcsec and $1.4 \cdot 10^{-18}$\sbunits, with data from \citet{Matsuda04} and \citet{Matsuda11}), while the lower gray area shows our fiducial choices for \Lya filaments (LAFs). Objects encircled in black show LAFs (LABs) with length $>400$\,pkpc ($>100$\,pkpc). Typically, the largest structures are also some of the most elongated ones.
}
\label{fig:filamentdetection_distribution}
\end{figure}

In Figure~\ref{fig:filamentdetection_distribution} we quantify filaments by their linear extent (size), area and surface brightness. The upper panel shows a histogram (i.e.\ space volume density) of filaments as a function of size, for different surface brightness thresholds. Overall, small structures are much more common than larger ones. Similarly, dim structures are much more common than brighter ones. The abundance of filaments with lengths below ${\sim}400$\,pkpc grows by a factor of a few when decreasing the surface brightness threshold by a factor of 10. For larger filaments, however, the number count is independent of surface brightness threshold, at and below our fiducial value of $10^{-20}$\sbunits. For $10^{-22}$\sbunits the distribution drastically changes its behavior as many of the larger structures merge. Finally, for smaller filament sizes from ${\sim}100-1000$\,pkpc, the decrease in number counts roughly follows a power-law with a slope parameter $\alpha>-2.0$ indicating larger filaments cover a comparable or larger sky fraction relative to smaller filaments.

The middle panel of Figure~\ref{fig:filamentdetection_distribution} shows a scatter plot of average filament surface brightness versus area. Each filament is represented by a single circle, where color indicates its circularity parameter. For detected structures above $100$\,arcsec$^{2}$, the number of objects drops significantly above a few times $10^{-19}$\sbunits, while the largest structures occur around this value. The number of filaments quickly drops towards the detection threshold of $10^{-20}$\sbunits, implying that low surface brightness filaments exist around brighter objects rather than on their own. With increasing area, the circularity decreases, demonstrating that the largest \Lya structures are elongated and filamentary in nature. On average, objects have increasingly circular shapes at larger surface brightnesses, indicative of halo-centered emission such as \Lya halos (LAHs), rather than intergalactic filaments.

To offer a face-value comparison, we also show the observational data points of the filament detections from \citet{Bacon21}, including only confident detections. We emphasize that the filament identification and measurement methods in that work differ from ours, and we do not intend a quantitative comparison at this stage. Each data point is plotted twice: once with the surface brightness and area at the redshift of its actual detection, and again with a second point representing the same filament at redshift $z=2$. To do so, we assume that the physical size and luminosity are both constant, which increases the surface brightness while decreasing its angular area.

Qualitatively, the confident detections in \citet{Bacon21} scaled to $z=2$ represent the brightest filaments in our simulations. These observed filaments have typical sizes around ${\sim}1000$\,arcsec$^2$, at the upper end of sizes that can be expected given the limited field of view. In the simulations, the majority of objects are smaller, and therefore harder to detect. In addition, observational searches within known LAE overdensity regions will bias the filament size distribution towards higher values, as we find to be the case in our models.

The lower panel of Figure~\ref{fig:filamentdetection_distribution} shows a scatter plot of filament luminosity versus area. While this plot is a simple mapping from the data shown in the middle panel, it offers complementary insights. First, we note that a relation between \Lya luminosity and area is unavoidable. Shaded regions show the allowed space for objects extracted for a given surface brightness threshold and smoothing scale. The upper limit is set by the minimum area for a smoothed point source at the surface brightness threshold. The lower limit is set by the minimum luminosity covered by an area above the surface brightness threshold.

We show to shaded regions for LABs and LAFs, defined by different surface brightness thresholds and smoothing scales. In each region, circles show detected \Lya structures at the respective SB threshold and smoothing scale. Just as in previous panel, we color the filaments by their circularity. Additionally, we show circles with a black edge color for LAB detections with $\geq 100$\,pkpc and for LAF detections with $\geq 400$\,pkpc.

The upper shaded region uses a surface brightness threshold of ${\left(1.0+2.0\right)/\left(1.0+3.0\right)}^{-4}1.4\cdot 10^{-18}$\sbunits and a Gaussian smoothing with FWHM of $1.4$\,arcsec as in the observational LAB datasets in~\citet{Matsuda04, Matsuda11} at $z=3$. In this case, extended \Lya structures in TNG50 have roughly similar characteristics, although we leave a detailed exploration of the abundance and properties of Lyman-alpha blobs for future work.

The lower shaded region is based on the values of our fiducial definition for extended LAFs. We find a power-law scaling of slope $1.1$ for the luminosity as a function of area for LAFs, which is in excess of the expected $1.0$ by construction. The typical surface brightness values of LAFs are roughly an order of magnitude above the surface brightness threshold for filaments with large areas. The largest filaments in TNG50 at $z=2$ have an area of $\sim 10^7$\,pkpc$^2$ and luminosities of a few times $10^{44}\,\rm{erg s^{-1}}$.

\begin{figure}
\centering
\includegraphics[width=0.47\textwidth]{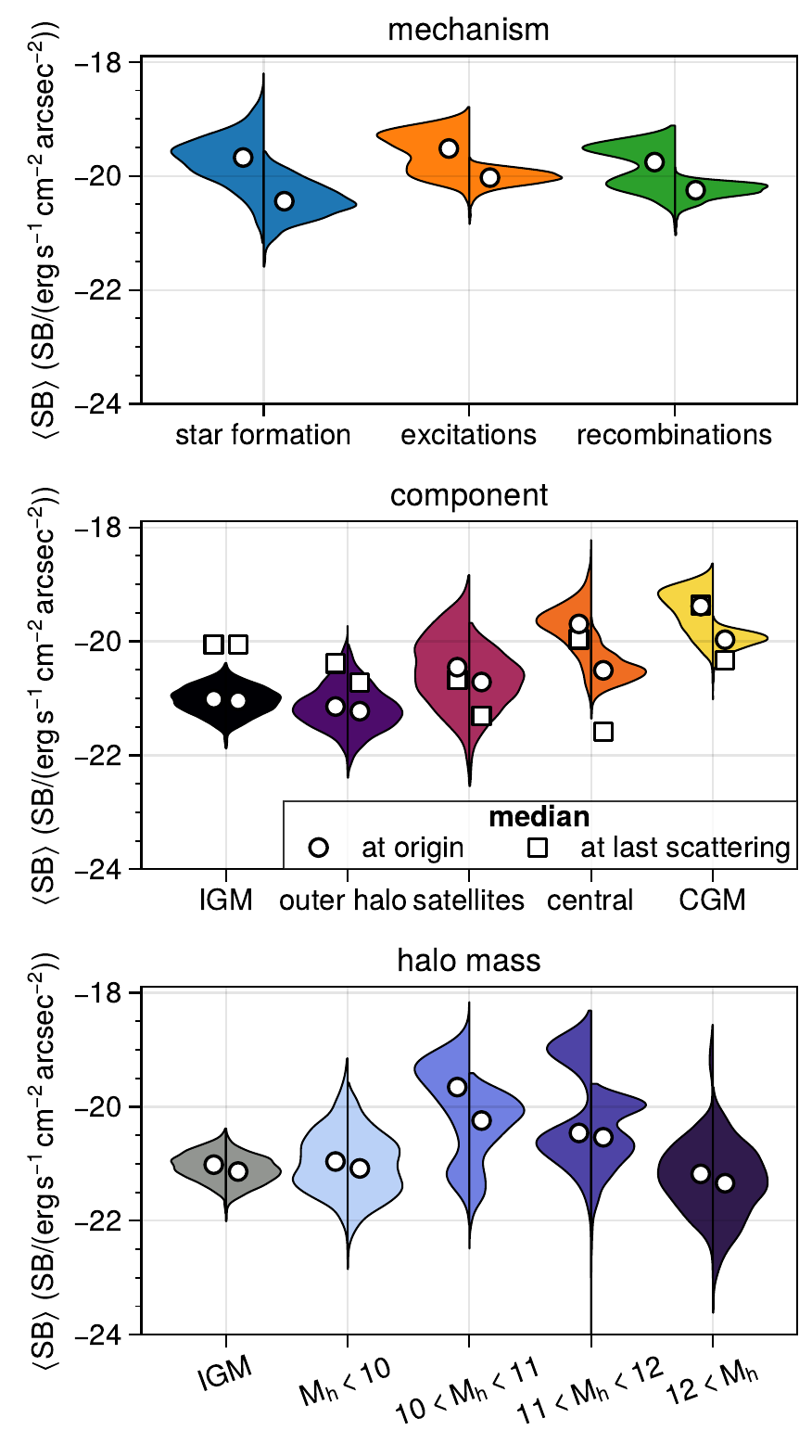}
\caption{Violin plots of the average \Lya surface brightness for detected filaments with $L>100$\,pkpc. We decompose this emission based on emission mechanism (top panel), spatial origin (middle panel), and contributing dark matter halo mass (bottom panel) at the point of emission. The left side of each violin shows the surface brightness distribution of detected filaments, for the given mechanism/component/halo mass. On the right side, the distribution is shown after masking out SB values above $10^{-19}$\sbunits. White circles indicate the median across all filaments. For the middle panel, we also show white squares indicating the median contribution of the spatial component photons scatter from last. We find a large variation in surface brightness for different contributors, but in general filaments are mostly powered by \Lya emission from star-formation and excitations, originating from the CGM and central galaxies of dark matter halos with masses of $10^{10}$ to $10^{11}$\,M$_{\odot}$.}
\label{fig:filamentdetection_props}
\end{figure}

In Figure~\ref{fig:filamentdetection_props} we show the mean surface brightness distributions of detected filaments with $L>100$\,pkpc as a number of violin plots. We split by emission mechanism (top panel), spatial component (middle panel) and halo mass (bottom panel). The left half of each colored regions shows the probability density function of the mean surface brightness within the defining contour at a surface brightness threshold of $10^{-20}$\sbunits. The right half of each colored region shows the mean surface brightness of the diffuse parts of the filaments, defined as contributions at SB values below $10^{-19}$\sbunits. The circles indicate the respective median values. In the panel for the spatial origin, we also show the median for the contributions at last scattering as squares.

Overall, we find that emission from stars and collisions contribute equally, with a substantial contribution from recombinations. For the diffuse regions (right side of violins), star-formation plays a subdominant role and collisions continue to dominate. We note that there is a large filament-to-filament variation in the relative contribution from stars, depending on the existence of nearby dust-poor massive galaxies. Emission from the CGM and central galaxies dominate in filaments (middle panel). Even in diffuse low surface brightness regions, the emission originating in the CGM remains the dominant contributor. Finally, halo masses between $10^{10}$ and $10^{11}$ are the most important in terms of contributing to filaments, also in their more diffuse regions (bottom panel).

While not explicitly shown, we also find that the relative importance of different contributions does not significantly depend on the filament area or length. In addition, if we decompose the total \Lya luminosity of detected filaments, rather than surface brightness, we obtain consistent results with those described above. Namely, for both small and large filaments -- independent of angular area -- these structures are dominated by \Lya photons sourced by collisional excitations (the other two mechanisms not far behind), which originate predominantly ($>$50\%) in the circumgalactic medium of intermediate-mass ($11 < \log{(M_{\rm halo} / \rm{M}_\odot)} < 12$) halos before scattering into filaments and towards the observer.

\begin{figure}
\centering
\includegraphics[width=0.47\textwidth]{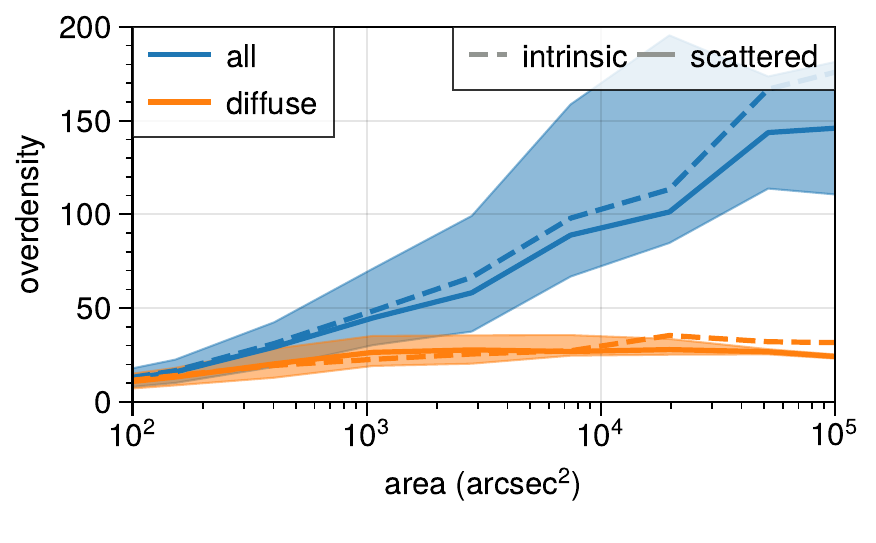}
  \caption{The luminosity-weighted dark matter overdensity that filaments trace as a function of filament area. The dark matter overdensity is computed using a $\sigma=100$\,pkpc Gaussian smoothing kernel. Lines show median relations, while shaded regions enclose the 16th and 84th percentiles. Blue plots overdensity for all filaments, while orange includes their diffuse (low surface brightness) regions only. Line style separately shows the result for the intrinsic (dashed) and scattered (solid) photons. Diffuse, large-scale \Lya filaments characteristically trace overdensities of order ten to twenty.}
\label{fig:filamentdetection_overdensity}
\end{figure}

In Figure~\ref{fig:filamentdetection_overdensity} we show the dark matter overdensity traced by \Lya filaments, as a function of their area. In blue we consider the entirety of each filament, while in orange we restrict to the diffuse, i.e.\ low surface brightness $<10^{-19}$\sbunits regions of filaments, thereby avoiding bright, halo-centric emission which may be embedded within them. We also show intrinsic (dashed) and scattered (solid) photons separately, to assess radiative transfer effects.

Our main finding is that \Lya filaments trace out increasingly overdense regions as their sizes increase. This relation increases monotonically, with overdensity rising from ${\sim}20$ to ${\sim}150$ as LAF area increases from $10^{2}$ to $10^{5}\,$arcsec$^{2}$. In contrast, the overdensity traced by the diffuse component is comparably constant, between $\sim 20$ and $\sim 50$. For filaments in their entirety, the corresponding overdensity is higher for intrinsic emission, as emission from the most massive objects escapes. For the diffuse component, the overdensity traced by intrinsic photons is lower by up to a factor of $2$, as photons from compact sources scatter into more diffuse regions.

Diffuse \Lya filaments, across a range of size scales, are an excellent tracer of the underlying matter of the cosmic web.

\section{Discussion}
\label{sec:discussion}

\subsection{The Lyman-alpha cosmic web and its physical origin}
\label{sec:dicussCW}

Our synthesized \Lya surface brightness maps enable us to assess the origin, and illumination, of \Lya emitting gas across a large cosmological volume at $z=2$. Our emission model self-consistently captures all major scenarios invoked to explain extended \Lya emission: scattered light from central sources, gravitational cooling, satellite galaxies, fluorescence from AGN, and the recent suggestion of contributions from unresolved LAEs \citep{Bacon21}.

Measuring isophotal areas above a fiducial surface brightness threshold $\rm{SB} > 10^{-20}$\sbunits and an imposed Gaussian PSF with a FWHM of $3.5$\,arcsec, we identify a range of \Lya emitting structures. At these fiducial values, large structures with $L\geq 400$\,pkpc, which we call \Lya filaments (LAFs), are increasingly more elongated and less round in shape. Our careful tracking of \Lya photons in terms of their emission mechanism, spatial origin, and originating dark matter halo mass lets us evaluate which scenarios contribute to the emergence of a ``\Lya cosmic web''.

\subsubsection{What powers Lyman-alpha filaments?}

On average, \Lya filaments (LAFs) source roughly half of their global emission from collisional excitation, while $30$\% is sourced by radiation from stars, leaving $20$\% from recombinations. Our results suggest that emission originating from cold gas outside of halos, i.e.\ from cold filaments of the intergalactic medium, does not significantly contribute to observable LAFs. Instead, most of the LAF emission arises from the CGM of halos, together with central galaxies and satellites. Even when considering the low surface brightness component ($<10^{-19}$\sbunits) of our filaments, the vast majority of emission stems from within gravitationally collapsed halos. Satellites provide between $10-20$\% of the luminosity and are a visible though minor contributor. Fluorescence from AGN, which is captured (in a simplified manner) within our model due to the inclusion of AGN radiative effects in the underlying TNG model, is likewise only a minor player. Instead, the CGM surrounding central galaxies is the most important spatial origin of \Lya photons for most filaments. In these denser environments, emission from collisions in the cold gas dominates. These photons, however, do not reach us directly -- $50$ to $80$\% of this emission reaches us only after scattering in the IGM, i.e.\ in filaments, causing them to shine in \Lya.

We can quantify the importance of these radiative transfer effects by defining a `boost factor', as the ratio between scattered versus intrinsic photon luminosity, for a given spatial component (not explicitly shown). If we do so, the IGM has a boost factor of $~1$\,dex, and the CGM itself is boosted below $10^{10}$\msun by a factor of a few, due to contributions from higher mass objects \citep[see][]{Byrohl21}. On the other hand, satellites and centrals have a net loss of luminosity with boost factors less than unity (${\sim}0.5 - 0.7$ on average), regardless of halo mass, representing intrinsic emission which escapes into the CGM and beyond. We now explore this scenario in more detail.

Figure~\ref{fig:filamentdetection_props} shows that halos with total mass between $10^{10}$\msun and $10^{11}$\msun source most filament luminosity. Contributions from fainter emitters with $L<10^{41}$\ergs, even in filament regions of low surface brightness, only play a minor role. This relative unimportance of low luminosity sources is due to three reasons.

First, our emission model has a steep $L_{\mathrm{Ly}\alpha}(M_{\mathrm{halo}})$ relation at the faint end, which reduces the slope of the luminosity function (where no observational constraints are available). Second, regardless of emission model, low luminosity emitters simply cannot contribute significantly, due to the low clustering of faint galaxies within \Lya filaments. To quantify this, we can model the surface brightness stemming from the LF only, i.e.\ ignoring environmental effects, as
\begin{align}
  \label{eq:integrateschechter}
  \mathrm{SB}_{\mathrm{LF}}\left(<L_\mathrm{det}\right) = \frac{\mathrm{d}_{\mathrm{A}}^{2}}{4\pi \mathrm{d}_{\mathrm{L}}^{2}} \Delta\mathrm{z} \int_{0}^{L_{\mathrm{det}}}\left(\delta_{\mathrm{fil}}\right(\mathrm{L}) + 1) L \Phi(L) dL.
\end{align}
Here, $\mathrm{d}_{\mathrm{L}}$ and $\mathrm{d}_{\mathrm{A}}$ are the luminosity and angular diameter distance, $\Delta\mathrm{z}$ is the slice depth in terms of physical length, and $\delta_{\mathrm{fil}}$ is the overdensity of emitters in a given filament compared to their cosmic mean. Using the Schechter form with parameters $\Phi_{\star}=6.32 \cdot 10^{-4}$\,cMpc$^{-3}$, $L_{\star}=5.29 \cdot 10^{42}$\ergs and $\alpha=-1.8$ from~\citet{Konno16}, faint emitters below $10^{41.75}$\ergs contribute up to $68\%$ of the total luminosity budget.

However, this assumes that the clustering, and thus the overdensity $\delta_{\mathrm{fil}}$, is constant across the luminosity range. This is not correct, as the clustering of halos is strongly mass dependent. \Lya filaments most commonly occur around more massive objects, and we find a median overdensity of $30.7$ for dark matter halos with total mass $10^{11}$\msun$\leq M_\mathrm{halo}\leq 10^{12}$\msun in filaments with lengths above $400$\,pkpc. The overdensity drops to $8.5$ and $3.7$ for mass ranges lower by $1$ and $2$\,dex respectively. Incorporating clustering, we find that halos with luminosities below $10^{41}$\ergs ($10^{40}$\ergs, $10^{39}$\ergs) contribute at most $22$\% ($7$\%, $3$\%), rather than $49$\% ($31$\%, $19$\%) when assuming equal clustering for all masses. Simply put, the abundance and distribution of low-mass halos (i.e.\ LAEs) prevents them from playing an important role in large-scale \Lya filaments.\footnote{Note that we calculate LAE overdensities in filaments based on our fiducial \Lya filament contours, i.e.\ the calculation is not fully self-consistent.}

Third and finally, the luminosity function captures only a fraction of the total \Lya emission in the Universe. While emission from the IGM itself is indeed negligible (see Figure~\ref{fig:mechs_components_overview}), emission from outside of observationally accessible apertures, i.e.\ below observational surface brightness limits, together with scattering from within these same apertures, out into the surroundings, contributes most of the global \Lya emission.

We turn to the recent observational detection of \Lya filaments of \citet{Bacon21}. That work identifies five filamentary \Lya structures at $z \sim 3$ and $z \sim 4.5$ at high significance. The authors conclude that most of the emission cannot be accounted for by the UV background, nor by detected \Lya emitters. Instead, they infer that very faint and unresolved \Lya emitters, down to $10^{38}$-$10^{40}$\ergs, are the major source of photons in the observed \Lya filaments. 

A full comparison with the findings of \citet{Bacon21} is not feasible, due to differences in methodology as well as redshift. Principally, this is due to the complexity of the filament identification algorithm in comparison to our more simple surface brightness threshold approach. With that caveat, our modeling does not appear to require any faint LAE population, i.e.\ emitters which are not simply already present in TNG50, to explain the existence and abundance of \Lya filaments. In contrast to their interpretation, our work instead suggests that rescattered contributions from resolved halos, particularly in the mass range of $10^{10}$ and $10^{11}$\msun, above $\mathrm{L}=10^{40}$\ergs produce significant numbers of \Lya filaments. However, we do agree with the conclusion that most of the filament emission does not stem from unbound IGM gas but instead from the CGM and, to a lesser degree, other halo-bound gas such as the satellites and from central galaxies themselves.

\subsubsection{How can we constrain different contributions in the Lyman-alpha cosmic web?}

Our analysis is based on the combination of the underlying TNG50 hydrodynamical simulation, the \Lya emission model, and the radiative transfer process. To disentangle these potentially degenerate effects, e.g.\ photon emission mechanisms, contributions by halo mass, the impact of radiative transfer, and the role of supernova and supermassive black hole feedback, we need observational measures which are separably sensitive to each of these aspects.

We have undertaken a number of variation runs, including switching each individual emission mechanism on and off, and turning the radiative transfer on and off. In each case, we run the entire calibration procedure from scratch, in order to match the observed LAE LF at $z=2$. Overall, we find that these experiments all produce a similar, and thus robust, space number density of $L>400$\,pkpc filaments. This suggests that our primary predictions, related to the properties and abundance of \Lya filaments are fairly robust. As a downside, however, degeneracies seem to be present which preclude the ability to clearly differentiate between many of the underlying physical processes. In the future, we plan to therefore explore:

\begin{enumerate}
    \item Alternative filament identification and characterization methods, such as multi-scale filtering and two-point statistics, or dedicated filament detection algorithms such as Disperse~\citep{Sousbie11}.
    \item Correlations with complementary \Lya data in filaments, particularly embedded LAE populations, including emitter luminosities, radial profiles, equivalent widths, and spectra.
    \item Correlations with complementary galaxy properties, particularly regarding AGN activity, stellar populations, and dust content.
\end{enumerate}

\begin{figure}
\centering
\includegraphics[width=0.47\textwidth]{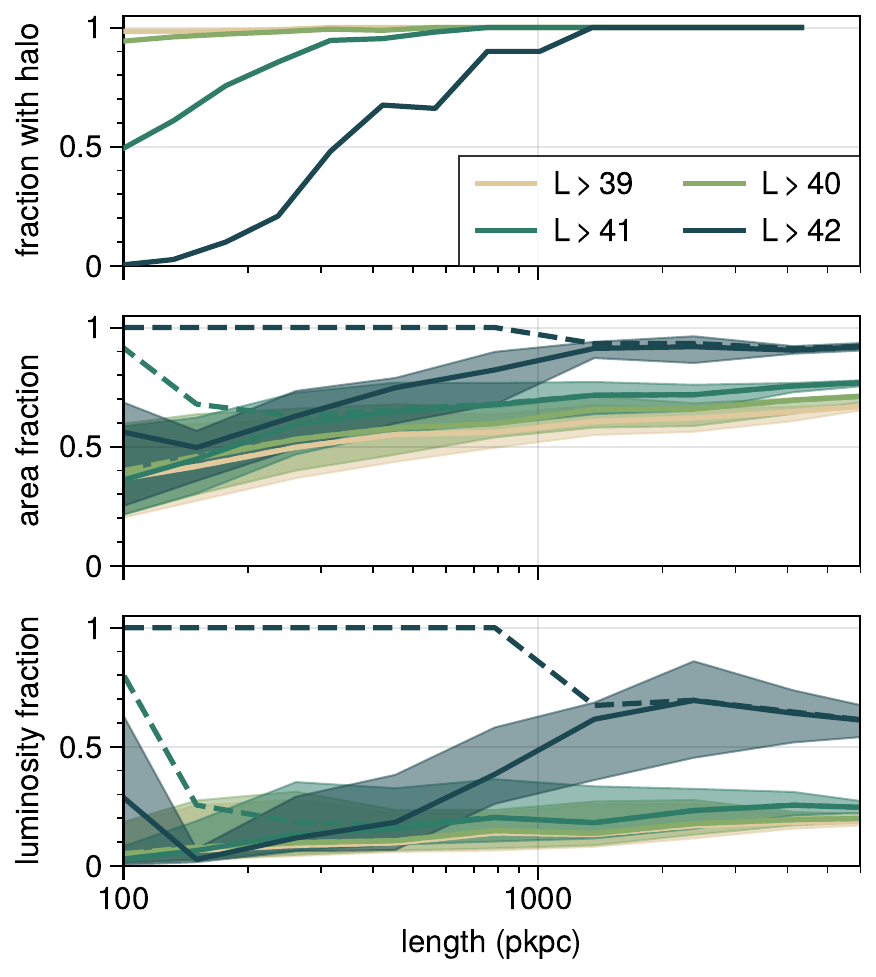}
  \caption{The fraction of filaments containing a \Lya-emitter (LAE) above a given luminosity (top panel). The fraction of filament area which remains after excluding (masking) all LAEs above a given luminosity (middle panel), and similarly for the filament luminosity fraction (bottom). All quantities are shown as a function of filament length. LAE luminosities are calculated as usual, while masked areas and luminosities are always evaluated as the total area and luminosity within the projected virial radius of each emitter which overlaps with the filament. Solid lines show medians for filaments with at least one emitter above the indicated luminosity thresholds (in $\log_{10}(\mathrm{L}/(\mathrm{erg}\,\mathrm{s}^{-1}))$), while dashed lines include filaments with no such emitters. While the largest filaments host bright LAEs, smaller filaments do not. Essentially all filaments contain at least one emitter above $10^{39}$\ergs, but the degree to which the area and luminosity of filaments is composed of emitters themselves depends on filament size (see text).}
\label{fig:LAEfilamentrelation}
\end{figure}

As a first step, we study the relation between LAEs and \Lya filaments in Figure~\ref{fig:LAEfilamentrelation} for our fiducial model. In the top panel, we show the fraction of filaments with a \Lya emitter above a given luminosity $\mathrm{L}$, denoting $\log_{10}(\mathrm{L}/(\mathrm{erg}\,\mathrm{s}^{-1}))$. \Lya luminosities are calculated as in our fiducial model with an aperture radius of $1.5$ arcseconds, see Section~\ref{sec:methodology}. We find that all filaments have at least one LAE above $10^{39}$\ergs, and the vast majority also have at least one LAE above $L>10^{40}$\ergs. However, at a luminosity threshold of $10^{41}$\ergs ($10^{42}$\ergs), only half of filaments with length above $100$\,pkpc ($300$\,pkpc) host such an LAE. With increasing length, filaments are more and more likely to contain such a bright emitter, while smaller filaments rarely do so.

In the middle and bottom panels, we investigate the degree to which filaments are actually made up of emitters. The middle panel shows the fraction of filament area area which remains after masking all LAEs above a given luminosity threshold, while the bottom panel similarly shows the fraction of filament luminosity. Solid lines give the median for filaments with at least one emitter above the luminosity threshold, while dashed lines show the median for filaments without any such emitter. Shaded regions show the associated central 68\% scatter.\footnote{To mask the projected area of each emitter, we use the area of the circular virial radius aperture which overlaps with the filament contour. For the luminosity masking, the photon contributions within the masked area are excluded. Note that the masked luminosity thus differs from the luminosity of the masked emitter, and no smoothing has been imposed in this step.} Considering bright emitters of $L>10^{42}$\ergs, we find that they account for the bulk of the \Lya luminosity for smaller filaments, and up to $40$\% for the largest. In terms of area, these emitters are responsible for up to half of the area for smaller filaments, but only $15$\% for the largest. If we instead consider LAEs down to a luminosity threshold of $10^{41}$\ergs, they represent a similar $\sim 50$\% of the area of small filaments, and $\sim 30$\% of the area of the largest filaments, as well as the majority ($\gtrsim 80$\%) of filament luminosity.

Going to even lower LAE luminosities does not appreciably change these results. That is, fainter emitters with $L < 10^{40}$\ergs do not contribute to \Lya filaments, in terms of either area or luminosity. For large filaments, more than half of the area and $15$\% of the luminosity remains unassociated with emitters $L>10^{39}$\ergs. This is consistent with the true fraction of $15$\% of \Lya luminosity within filaments reaching us from unbound gas. Further, only $10$\% of this initial $15\%$ luminosity fraction originates within the IGM. As previously found, the surface brightness level of the diffuse IGM within filaments is thus boosted by an order of magnitude through scattered photons from nearby halos compared to its intrinsic emission.

In addition to our fiducial virial radii mask, we have also considered a fixed 3 arcsecond aperture for all emitters (not shown), which is more accessible observationally. We find that more than $90$\% of filament area and $60$\% of filament luminosity remain after masking above a luminosity threshold of $10^{41}$\ergs. Decreasing the emitter luminosity threshold removes an increasing fraction of area and luminosity, as opposed to the plateau effect seen for the virial radii apertures. This is, however, driven by the substantial amount of IGM masked. For $L>10^{39}$\ergs, the luminosity and area fractions remain above those of the virial radius aperture masking.

\subsubsection{Connection with Lyman-alpha halo radial profiles}

\begin{figure}
\centering
\includegraphics[width=0.46\textwidth]{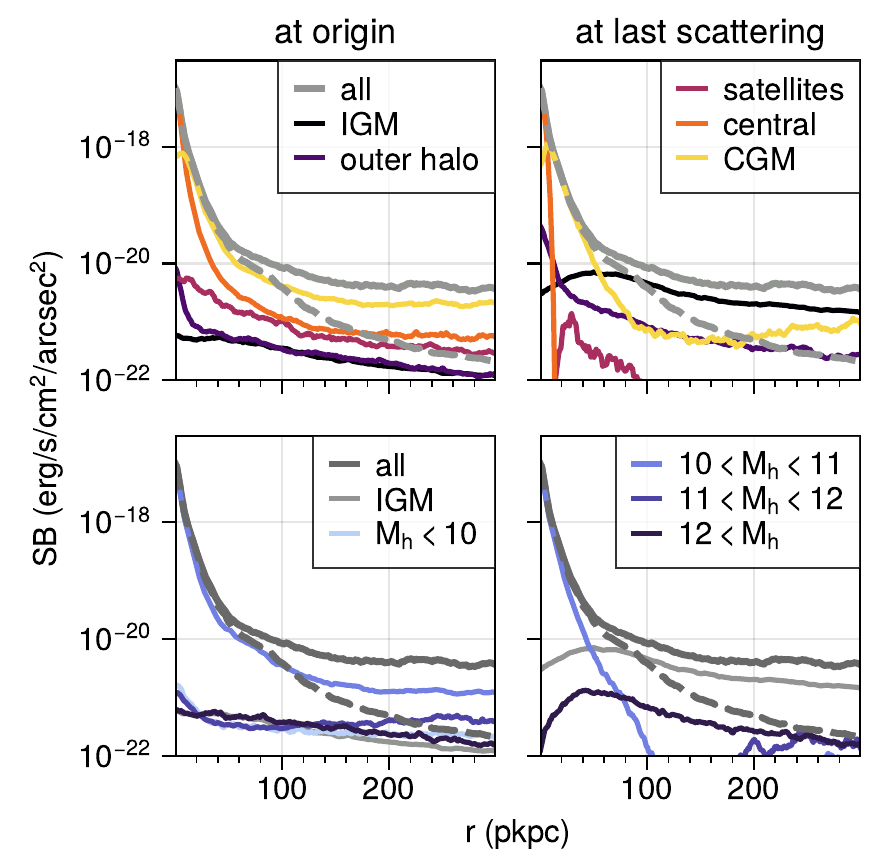}
  \caption{Median radial profiles of extended \Lya emission centered on galaxies, i.e.\ \Lya halos \citep[following][]{Byrohl21}. We include and stack LAEs with REW$>20$\angstrom between $L=10^{42}$ and $10^{43}$\ergs, adding a Gaussian PSF (FWHM = $0.7$\,arcsec). In the top panels, we decompose radial profiles by their spatial component. In the bottom panels, we decompose the radial profiles by their sourcing halo mass. On the left, we take scattered \Lya photons and decompose based on the component at their spatial location of emission. On the right, we instead decompose the scattered \Lya photons at the location of their last scattering. The median profiles quickly decrease from their central peak of ${\sim}10^{-17}$\sbunits, and flatten around ${\sim}50$\,pkpc at $\sim\cdot 10^{-20}$\sbunits. The dashed line indicates the contribution from the targeted halo alone, demonstrating that this flattening is caused by scattered photons originating in other, nearby halos. In fact, the flattened profiles are dominated by emission from the CGM of halos with masses between $10^{10}$ and $10^{11}$\msun scattering into the IGM.}
\label{fig:rps}
\end{figure}

While the detection of diffuse \Lya filaments remains challenging, observations can increasingly identify brighter emission centered on galaxies, in the form of \Lya halos and \Lya blobs \citep{Steidel11,Momose16,Borisova16,Leclercq17,Cai19}. Their stacked radial profiles are observed to flatten at large radii in HETDEX data, far beyond the halo boundary \citep{LujanNiemeyer22}, a phenomenon identified in our previous analysis of TNG50 \citep{Byrohl21}. We now connect these halo-centric profiles to our current cosmic web study.

Figure~\ref{fig:rps} shows the median stacked radial profile of REW$>20$\angstrom LAEs with luminosities between $10^{42}$ and $10^{43}$\ergs. We show the contributions split by the spatial component they last scatter from, i.e.\ observable \Lya photons (left) and intrinsically originate from (right). We see that the large distance flattening primarily arises due to contributions last scattered by the IGM\@. However, these photons mainly originate in the CGM, with smaller contributions by central galaxies and centrals. This boosts the IGM luminosity by roughly an order of magnitude over its intrinsic emission, consistent with our findings for the diffuse parts of \Lya filaments.

The original source of these photons in the outskirts of \Lya halo profiles, as shown in the bottom panels, are intermediate mass halos with $10^{10}$\msun$\leq M_\mathrm{halo}\leq 10^{11}$\msun. Even more massive halos start to non-negligibly contribute at large, $\gtrsim$ 100s of kpc, distances \citep[see also][]{Byrohl21}. Our current work demonstrates that at these SB values, \Lya halo radial profiles in fact trace filaments of the cosmic web itself. In particular, they begin to represent contributions from the fuzzy outskirts of filaments illuminated by scattered photons.

\subsubsection{Lyman-alpha luminosity global budget}
\label{sec:discussobs}

The luminosity density of \Lya photons gives us a handle on the global emissivity across a large-scale, cosmological volume. The luminosity function of Figure~\ref{fig:lf} has a total luminosity density of $\dot{\rho}_{\Lya}=4.3 \cdot 10^{39}$\ldens when integrated across the observed luminosity range $\mathrm{L}>10^{41.75}$\ergs. By construction, this is close to the integrated Schechter form from~\citet{Konno16} within ${\sim}10\%$. However, this is an order of magnitude lower than the true global luminosity density of $\dot{\rho}_{\Lya}=4.4 \cdot 10^{40}$\ldens. For \Lya emission included within the luminosity function, the majority of emission stems from faint sources. In particular, \Lya emitters below $10^{41.75}$\ergs contribute $\dot{\rho}_{\Lya}=6.1 \cdot 10^{39}$\ldens. The majority, $5.5\cdot 10^{39}$\ldens, of this stems from emitters above $10^{40}$\ergs. Furthermore, the observational \Lya LF is constructed by including only emitters with an equivalent width cut of REW$>20$\angstrom. Our fiducial model shows that this REW threshold removes \Lya emitters with a total luminosity density of $\dot{\rho}_{\Lya}=2.8 \cdot 10^{39}$\ldens. If we instead consider all \Lya emitters, keeping fixed the $1.5$\,arcsecond aperture, they account for ${\sim}30\%$ of the global \Lya luminosity.

Of the ${\sim}70$\% of the global \Lya budget not captured within this aperture\footnote{We compare against~\citet{Konno16} with an aperture radius of $1.0-1.5$\,arcsec. This corresponds to $8.6$ to $12.9$\,pkpc at $z=2.0$, corresponding to the virial radii of halos with total mass up to $2\cdot 10^{9}$\msun.}, $60$\% reach us from bound halo gas outside of the aperture, and $40$\% from the IGM. Again, we stress the importance of \Lya radiative transfer for this result. Intrinsically, without radiative transfer, only $2$\% of the global budget originates in the IGM, but this value is boosted to $27$\% after scattering. Similarly, intrinsic emission from within the aperture would account for ${\sim}50\%$ of the total budget, but a substantial fraction of these photons are redistributed outside the aperture after radiative transfer.

In agreement with this finding, \Lya intensity mapping results at $z\sim 2$ have found a significantly larger \Lya photon budget than accounted for by detected LAEs~\citep{Croft16, Croft18}. However, we caution that since these \Lya intensity mapping experiments are constructed using quasars for their cross-correlation analysis, their large \Lya luminosity density estimate might result from the biased quasar environment and proximity effects. Nonetheless, a recent analysis by~\citet{Lin22} yields $\dot{\rho}_{\Lya}=6.6^{+3.3}_{-3.1} \cdot 10^{40}$\ldens, and the authors argue that \Lya emission from star-forming galaxies not associated with resolved LAEs could be responsible for such a high luminosity density. This observational estimate of the luminosity density is 50\% higher than our fiducial result, but consistent to better than $1\sigma$.

\Lya filaments with linear extents above $100$\,pkpc ($400$\,pkpc, $1000$\,pkpc) contain $76$\% ($59$\%, $47$\%) of the global \Lya photon budget, a substantial amount compared with the \Lya luminosity captured by the LAE luminosity function.

With respect to the brightest luminosity \Lya sources, the observed luminosity function above $L > 10^{43} \rm{erg s^{-1}}$ flattens considerably, and these rare, bright emitters are directly related to AGN~\citep{Konno16}. Our model does not include the intrinsic \Lya emission from AGN, which are regardless quite rare in the relatively small volume of TNG50. As a result, we underestimate the abundance of LAEs above this threshold. However, only a small fraction of the total luminosity density arises from these sources.

\subsection{Prospects of observing the Lyman-alpha cosmic web}
\label{sec:prospects}

Our modeling provides quantitative predictions for the expected occurrence of observable \Lya filaments (LAFs). For filaments of a linear size $L>400$\,pkpc, we predict a moderate density of $2 \cdot 10^{-3}$\,cMpc$^{-3}$ above a surface brightness threshold of $10^{-20}$\sbunits, for a line-of-sight slice depth of $5.7$\angstrom and a smoothing with a FWHM of $3.5$\,arcsec (see Figure~\ref{fig:filamentdetection_distribution}).

MUSE has a $1$\,arcmin$^2$ footprint. This covers a volume of $1800$\,cMpc$^{3}$ ($3600$\,cMpc$^{3}$, $5300$\,cMpc$^{3}$) between $z=2.0$ and $2.5$ ($3.0$, $3.5$). Neglecting surface brightness dimming, redshift evolution, and noise complexities beyond a surface brightness threshold, this would imply an average of $2.8$ ($5.5$, $8.2$) filaments above a surface brightness threshold of $10^{-20}$\sbunits for \Lya structures with $L>400$\,pkpc. At a higher surface brightness threshold of $3 \cdot 10^{-20}$\sbunits ($10^{-19}$\sbunits) this number reduces to 1.4 (0.4) between $z=2.0$ and $2.5$.

Three other relevant IFU instruments are KCWI, BlueMUSE, and HET-VIRUS, with field of view footprint sizes $0.2$, $2.0$ and $54.2$ times the footprint of VLT-MUSE, respectively. The expected detection counts can be scaled linearly for a first estimate. Clearly, statistically significant samples of $\gtrsim 10$ filaments can only be achieved for large survey volumes. With current instruments, this will require many pointings and mosaicing, with the exception of HET-VIRUS.

\begin{figure}
\centering
\includegraphics[width=0.47\textwidth]{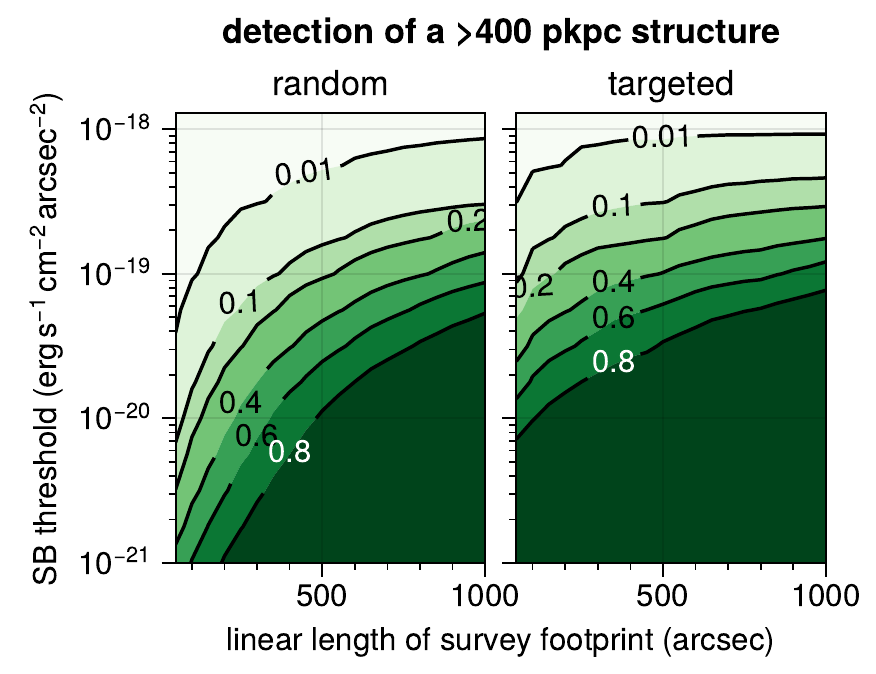}
  \caption{The probability for detecting a \Lya filament with a length above $400$\,pkpc, as a function of the total footprint size of a hypothetical survey, and its surface brightness threshold, based on our fiducial smoothing choice of a $3.5$\,arcsec FWHM\@. Lines show equiprobable contours of $1$, $10$, $20$, $40$, $60$ and $80$\% to detect at least one filament. The left panel shows the probabilities for random pointings, and the right panel shows the situation for targeted pointings centered on a random LAE with $\mathrm{L}>10^{42}$\ergs. As expected, a small field of view combined with a high SB threshold results in rare detections (upper left), while a large field of view combined with a sufficiently low SB threshold should easily detect multiple \Lya filaments (lower right). Any survey with large wavelength coverage, e.g. using an IFU instrument instead of a narrow-band imager, will have a search path-length much larger than this slice depth, which must be accounted for (see text).}
\label{fig:survey_probabilities}
\end{figure}

We further quantify the observability of \Lya filaments in Figure~\ref{fig:survey_probabilities}, which shows the expected number of detected \Lya filaments with $L\geq 400$\,pkpc for a hypothetical survey, as a function of survey sky footprint (linear length, assumed to be square) and surface brightness threshold. We again adopt a fiducial redshift slice depth of $5.7$\angstrom. Lines show contours for detection probability in a single redshift slice, between $1$\% and $80$\%. We consider two cases: random pointings (left panel) and targeted pointings centering on a previously known \Lya emitter with $L>10^{42}$\ergs (right panel).

For reference, a single pointing of VLT-MUSE corresponds to $60$ arcsec, which is the left edge of this figure, while a single pointing of HET-VIRUS corresponds to $951$\,arcsec (neglecting the non-unity filling factor), at the right edge of this figure. Regarding the sensitivity, for example, the 10-hour MUSE UDF data, with a 3x3 tiling (${\sim}180$ arcsec total linear extent), reaches a $1\sigma$ SB sensitivity of \mbox{$5.5 \cdot 10^{-20} \,\rm{erg\, s^{-1}\, cm^{-2}\, arcsec^{-2}}$\,\AA$^{-1}$} in a 1 arcsec aperture \citep{Leclercq17}.

For large footprints, the probabilities for random and targeted pointing strategies at $z=2$ are similar.\footnote{We have also focused our analysis exclusively on $z=2$ for simplicity. Significant evolution in the physical properties of \Lya filaments towards higher redshift would change any quantitative observability metrics beyond simple rescaling. Our RT post-processing can, however, be applied directly at higher redshifts in the simulations in future work.} We predict a $50$\% detection probability at $10^{-19}$\sbunits for a $1000$\,arcsec survey footprint. As these are achievable surface brightness levels, our modeling suggests that a sufficiently large survey at such a sensitivity will easily detect filaments of the \Lya cosmic web. The situation is more difficult for small survey footprints. At $60$ arcsec and the same SB threshold of $10^{-19}$\sbunits, the targeted pointing has only a $10$\% detection probability, while the random pointing has only a $1$\% chance. To achieve the same probability, the random pointing would need to reach a SB threshold roughly an order of magnitude lower, ${\sim}2\cdot 10^{-20}$\sbunits.

However, one promising avenue for the observational detection of diffuse \Lya filaments is with integral field spectrograph/unit (IFS/IFU) instruments. In this case, a significant wavelength coverage along the spectral dimension implies that the observed volume along the line-of-sight direction is significantly larger than our fiducial, narrow slice width of only $5.7$\angstrom. A rough estimate for the detection of filamentary structures in an extended survey volume along the line-of-sight can be estimated from the number densities given at the beginning of this section and in Figure~\ref{fig:filamentdetection_distribution}.

Our use of a single SB threshold number to characterize the sensitivity of an observation is only meant to provide intuition. The real power of the modeling approach presented here is that, for any proposed observational strategy, detection probabilities can be quantitatively assessed. Specific targeting and/or stacking concepts can be evaluated a priori \citep[for example, oriented pair experiments;][]{Gallego18}. The ideal observational experiment, given instrumental and practical constraints, can be designed to detect the cosmic web in \Lya emission. We can also test new data analysis/statistical methods to better identify and characterize such structures in difficult regimes. For example, in low signal-to-noise data, in the presence of substantial foreground contamination, and/or when extended low surface brightness emission could be lost due to overaggressive background subtraction on a limited field of view.

\subsection{Modeling uncertainties and future directions}
\label{sec:modeluncertainties}

The emission of \Lya photons and their radiative transfer are sensitive to the physical state of gas, from galactic, to circumgalactic, to intergalactic scales. We therefore discuss current limitations of our \Lya emission model, the \Lya radiative transfer method, and the underlying hydrodynamical simulation.

\subsubsection{Lyman-alpha emission modeling}
\label{sec:discussem}

Building a physically motivated and realistic emission model for \Lya on top of a hydrodynamical simulation result is challenging. Various \Lya emission mechanisms have been considered in the context of numerical studies~\citep{Kollmeier10, Cen13, Lake15, Smith19a, Mitchell21, Byrohl21}, although consensus on which mechanisms dominate for different observed classes of \Lya emitting objects has not been reached in the past. Cold dense gas accounts for the majority of the \Lya emission in our model, before scattering outwards into lower density surroundings. This implies that modeling the emission from overdense regions is always critically important. Because of radiative transfer effects, this is true even if one is primarily interested in lower density regions, such as cosmic web filaments, as we are here.

Characterizing the mean diffuse emission for collisions and recombinations within the virial radius as a function of halo mass (Figure~\ref{fig:halolums}, calibrated model), we find a steep relation with a power law slope $\alpha>-2$ below $10^{10}$\msun. There is a mild discontinuity around $10^{11}$\msun where halos with radiative feedback from AGN start dominating the sample. Compared to other models, our fiducial, calibrated model shows a similar mass dependence as the lower-bound model by~\citet{Faucher-Giguere10}. Our $L_{\mathrm Ly\alpha}(\mathrm{M}_{\rm{halo}})$ relation predicts moderate luminosities, between the conservative model of~\citet{Faucher-Giguere10} and brighter alternatives which have been previously applied to cosmological simulations~\citep{Yang06, Goerdt10, Rosdahl12}.\footnote{Note that all studies we compare against are at $z\sim 3$, with exception of~\citet{Yang06}. Extrapolating as $L\sim {(1+z)}^{1.3}$ ~\citep{Goerdt10} yields only a $30\%$ decrease in luminosity from $z=3$ to $z=2$, so we do not expect our result for diffuse emission to change significantly at $z=3$.}

Ionizing radiation from stellar populations dominates \Lya emission in star-forming regions. In our previous work~\citep{Byrohl21}, we adopted a simple model where \Lya emission was proportional to the local star-formation rate of \textit{gas}. Instead, the new model presented in this work is fundamentally different: we model emission for stellar populations according to their age and metallicity of \textit{stars}, and apply the calibration described in Section~\ref{sec:rescaling}. The empirically calibrated model accounts for dust, including the destruction of \Lya photons by dust, which would otherwise not be captured. This small-scale physics constitute a sub-resolution model, as it occurs below the resolution scale of the simulation. The previous lack of a dust model led to an overestimate for emission from star-forming regions in massive galaxies. Now, by calibrating the stellar luminosities against observed luminosity functions, we substantially limit the stellar \Lya photon budget. Figure~\ref{fig:laeprops} indicates that our rescaled model performs decently against other observables, and sufficiently well for our purposes. Future advancements could incorporate additional observables into the fitting procedure. This would help to break remaining degeneracies, and improve the overall physical fidelity of the emission model.

One such observable is the observed extent of \Lya halos. For example, \citet{Leclercq17} find an exponential scale length $r_{0}$ of $4.4\pm 1.5$\,pkpc at $z\sim 3$. With our fiducial model, for \Lya halos with $41.5\leq\log\left(\mathrm{L}/\mathrm{L}_{\star}\right)\leq 42.5$ and REW$\geq 20$\angstrom, we find an exponential scale length of $r_{0}=7.0_{-1.9}^{+3.7}$\,pkpc (see Figure~\ref{fig:rps}), which is ${\sim}60$\% larger than the observational profiles. However, as previously demonstrated in~\citet{Byrohl21}, the exponential scale lengths in TNG50 \Lya halos match observations when dust attenuation is negligible. At the same time, Figures~\ref{fig:lf} and~\ref{fig:laeprops} make clear that significant dust attenuation is needed to match the observed \Lya luminosity function at $z=2$. It is possible that a more sophisticated dust model could reconcile these differences, i.e.\ simultaneously matching the exponential scale lengths and the luminosity function. At present, this tension tentatively suggests that our model produces too much diffuse emission from collisions and recombinations, and has too little escaping \Lya luminosity from dense star-forming regions.

In the current approach, dust impacts only the stellar population sourced component, and does not modify any of the diffuse emission mechanisms. The physical motivation is that optical depths and dust destruction is substantially smaller away from the ISM\@. In our best-fit model, diffuse emission (i.e., recombinations and collisions) dominates the luminosity budget for halos with observed luminosities below $10^{40}$\ergs as star-formation ceases in the TNG50 simulation. However, there is an additional peak around $10^{42}$\ergs where diffuse emission dominates over stellar emission, potentially hinting at an overestimate in the diffuse emission.

As our dust model is based on the rescaling of intrinsic luminosities, it will naturally overestimate spectral as well as spatial redistribution, when compared to runs self-consistently incorporating dust~\citep[see e.g.][]{Laursen09}.
To explore this scenario, we test an alternative model which explicitly includes a dust description for diffuse emission. This leads to a reduction of $15\%$ and $18\%$ for recombinations and collisions, respectively, which would not change our overall conclusions.\footnote{To perform this test, we ran a \Lya radiative transfer simulation with the dust optical depth given by Equation~\ref{eq:uvdust}, an albedo $A=0.33$ and a Henyey-Greenstein phase function with a parameterization of $g=0.68$.} Even if star-forming regions contributed more than our fiducial model suggests, making them the dominant emission mechanism, our finding that compact spatial components and star-forming galaxies illuminate \Lya filaments through resonant scatterings would remain robust.

To bracket the uncertainties in the modeling of collisional excitations, we ran a simulation without any collisions. In this case, we can still find a rescaling model to match the observed luminosity function in agreement with the \Lya escape fractions and equivalent width distribution. We find that the majority of the global luminosity budget due to collisions in the fiducial model ($2.1\cdot 10^{40}$\ldens) is countered by an increased \Lya emission around stars in this variation run ($1.3\cdot 10^{40}$\ldens $\rightarrow$ $2.4\cdot 10^{40}$\ldens). This leads to an increased fraction of photons scattered into filaments originating from galaxies. Interestingly, the CGM still contributes $34\%$ of the global luminosity budget, primarily through recombinations but also stellar populations. We expect similarly robust conclusions when ignoring recombinations rather than collisions. The filament density for $L>400$\,pkpc decreases by ${\sim}20$\% ($40$\%, $80$\%) at a surface brightness threshold of $10^{-20}$\sbunits ($3\cdot 10^{-20}$\sbunits, $10^{-19}$\sbunits) compared to our fiducial model.

Using a multiscale approach, such as the starlet wavelength transform~\citep{Starck98} used in~\citet{Bacon21}, could offer complementary information to differentiate underlying contributions. In the future, purely \Lya observation-based statistics, such as filament number densities, shapes and presence of luminous LAEs, in addition to the UV luminosities of galaxies within filaments, can be explored to differentiate these scenarios (see Figures~\ref{fig:filamentdetection_distribution} and \ref{fig:LAEfilamentrelation}).

\subsubsection{Lyman-alpha radiative transfer}
\label{sec:discussrt}

\Lya filaments have recently been studied with hydrodynamic simulations in two past works which are particular relevant to our study~\citep{Elias20, Witstok21}. Neither of these studies incorporates radiative transfer or scattering, i.e. their results would correspond to our intrinsic emission. However, our analysis shows that the surface brightness of the \Lya cosmic web is considerably boosted by radiative transfer. This finding directly invalidates the assumption of previous work hypothesizing that scattering does not substantially affect, or even decreases, \Lya filament detectability.

For example, a consistency check of an individual filament in~\citet{Elias20} shows that \Lya radiative transfer redistributes emission to lower surface brightness, boosting the occurrence of spaxels with SB values below $10^{-21}$\sbunits (see their Figure A1). Such low SB values are observationally unreachable, leading to the idea that radiative transfer decreases detectability. Their model, however, only includes emission from gas outside of halos. When incorporating emission from within halos, which contributes $98$\% of our global emission, we find the occurrence of SB values below $10^{-19}$\sbunits, typical for \Lya filaments, to be boosted by radiative transfer (see our Figure~\ref{fig:sbpdf_mechs} and~\ref{fig:sbpdf_overdensity}). Similarly, \citet{Witstok21} argue that the impact of \Lya radiative transfer is negligible for the observability of \Lya filaments, however this assumption is not actually tested with radiative transfer simulations. Based on the lack of a \Lya radiative transfer treatment, and the lack of emission from within halos, we consider the expectations from~\citet{Elias20} and~\citet{Witstok21} to be lower limits for the purposes of \Lya filament detectability.

In particular, \citet{Witstok21} mitigate modeling uncertainties at high densities by imposing a overdensity threshold above which \Lya luminosity contributions are ignored. This limits the surface brightness from recombinations to the ``mirror limit'' at which an optically thick cloud reflects most of the UVB as \Lya emission. This would correspond to a value of $\sim 2.5\cdot 10^{-20}$\sbunits for the UVB of TNG50 at $z=2.0$. However, this assumption yields a very conservative lower limit for the recombinational (as well as collisional) \Lya emission in filaments, as it neglects any recombinations that can still be sourced in collisional equilibrium, as well as local sources and subsequent enhancements through \Lya scatterings. In our model, we do not adopt such an assumption.

On the other hand, \citet{Elias20} find that \Lya filaments are, in general, brighter and more detectable in simulations which incorporate the TNG galaxy formation model, in comparison to simpler ``no feedback'' simulations. The authors suggest that the ejection of gas from lower mass galaxies, due to stellar feedback-driven winds, can distribute more gas into filaments. They then speculate that the lack of detection of \Lya filaments to date suggests that this gas redistribution may be too strong in the TNG model. However, given the caveats discussed above, namely the importance of radiative transfer together with the lack of quantitative and/or statistical comparisons with observational data in~\citet{Elias20}, we do not find this suggestion compelling.

In fact, we conclude that our TNG50+RT predictions for \Lya filament abundance and detectability are roughly consistent with the recent available observations of \citet{Bacon21}. While we cannot directly compare results given the different filament identification method, our model suggests a low single digit count of extended \Lya filaments similar to the largest structure in~\citep[`group2' of][]{Bacon21}. The observed structure has an linear extent above $450$\,pkpc and a circularity below $0.2$, consistent with our expectations. Our simulated sample contains numerous smaller filaments, which are however not identified in~\citet{Bacon21}, by construction, as they focus on filaments around LAE overdensities.

Also of interest, \Lya blobs (LABs) represent another class of extended \Lya structures beside \Lya halos and \Lya filaments. The ${\sim}100$\,pkpc sizes of LABs fall between those of LAHs and LAFs, but their surface brightness threshold is the highest at ${\sim}10^{-18}$\sbunits~\citep{Matsuda04, Matsuda11}. LAHs, LABs and LAFs are primarily defined by selection in surface brightness and typical size. Nevertheless, they stem from the same underlying gas, and the same underlying emission mechanisms, albeit potentially in different configurations. When we consider the same surface brightness threshold and point spread function of \citet{Matsuda11}, we find one LAB with an area $>100$\,arcsec$^{2}$ and circularity of $0.33$ (see bottom panel of Figure~\ref{fig:filamentdetection_distribution}). This is broadly consistent with the observational occurrence rate and observed circularities~\citep{Yang10, Matsuda11}, although the small volume of TNG50 makes this comparison challenging.

\subsubsection{The TNG model and physical gas state}
\label{sec:gasstate}

With respect to underlying cosmological simulation, the main limitations of interest based on our use of TNG50 of the IllustrisTNG project are discussed in \citet{Byrohl21}, to which we refer the reader for more discussion.

The TNG simulations do not solve the equations of radiation-hydrodynamics (RHD), i.e.\ they do not directly incorporate radiation on-the-fly during the simulation. In the current context, this could be used to include currently missing local sources of ionizing radiation, namely stellar populations, as photoionization and photo-heating terms. While such simulations are increasingly feasible for in cosmological volumes down to $z\sim 5$ in the study of cosmic reionization \citep{Rosdahl18, Ocvirk20, Kannan22}, as well as zoom simulations of individual galaxies \citep{Rosdahl12, Mitchell21, Costa22}, they are currently not feasible for large cosmological volumes to low redshift. This is an absolute requirement for predicting the observability of the \Lya cosmic web at $z \sim 2-3$, which is why TNG50 and its combination of resolution and volume is our preferred tool.

While TNG50 is not a RHD simulation, it does incorporate simplified models of radiation in the most important regimes. Namely, the TNG model has an on-the-fly treatment for ionizing radiation from AGN, approximated as a spherically symmetric radiation field in the optically thin limit. Its impact on \Lya emission around galaxies is critically important \citep[as discussed in][]{Byrohl21}. Through photoheating and -ionization, AGN radiation significantly boosts emission from both recombinations and collisions (see Section~\ref{sec:sbdist}). As a result, for halos above $3\cdot 10^{10}$\msun, \Lya luminosities increase by a factor of $10$ for recombinations, and a factor of $2$ for collisions. For reference, $\sim 40$\% ($\sim 80$\%) of the gas mass (\Lya emission) in TNG50 at $z=2$ arises from gas experiencing a local AGN radiation field in addition to the UVB.

We note that TNG does not include a similar model for stellar populations, such that the ionizing radiation from \textit{local} stars is not included in our modeling. However, TNG does incorporate a spatially uniform, time variable metagalactic background radiation field \citep[UVB;][]{Faucher-Giguere09}, which is critically important for setting the physical state of low-density gas. In denser regions, gas can self-shield from this external radiation, and TNG includes a self-shielding model based on \citet{Rahmati13}, to reproduce the average ionizing field suppression. Collisional excitations are particularly sensitive to the gas state in the simulation, due to the exponential temperature dependence around ${\sim}10^{4}\,$K~\citep{Furlanetto05, Faucher-Giguere10}, as we discuss below.

While TNG50 is the highest resolution cosmological hydrodynamical simulation which exists at its volume \citep{Nelson19a}, the halo mass function is only resolved, if we require a minimum dark matter particle count of 1000, down to halo masses of $4\cdot 10^{8}$\msun, which corresponds to the median halo mass at an intrinsic \Lya luminosity of $2\cdot 10^{39}$\ergs for our fiducial (rescaled) model. In our model, \Lya luminosities are drawn from the averaged initial mass function, as star particles in the TNG model represent entire stellar populations. Particularly at the faint end, we thus do not incorporate the scatter introduced by sampling individual, massive stars that could potentially alter the \Lya emission in low-mass galaxies.

Despite the promising level of consistency between our model results and available \Lya observations, as discussed above, the feedback model of the TNG simulations, both in terms of expelled gas and as a photoionization source in the TNG model, is effective in nature and its details have significant theoretical uncertainty. Our methodology to predict \Lya emission across scales enables new empirical constraints on the underlying galaxy formation physics, through future, more rigorous comparisons with observations.

These constraints on the \Lya emitting gas probe scales, and phases, distinct to other observable tracers, thereby providing orthogonal tests of gas in the TNG simulations, beyond existing explorations, e.g. total gas fraction~\citep{Pillepich18, Terrazas20, Ramesh22}, absorption as well as emission from cool MgII~\citep{Nelson19a, Nelson21}, warm-hot OVI abundance~\cite{Nelson18a}, hot x-ray emission~\citep{Truong20, Truong21}, the Sunyaev-Zeldovich signal~\citep{Pop22}, and the large-scale (re)distribution of gas by feedback~\citep{Ayromlou22}.

The observables of \Lya, from interstellar medium to intergalactic medium scales, are fundamentally linked to the underlying hydrodynamical simulation and its baryonic feedback models.

\section{Conclusions}

In this paper, we develop a comprehensive theoretical model for \Lya emission across scales, from individual Lyman-alpha emitters to extended structures such as Lyman-alpha halos, blobs and filaments of the cosmic web. To do so we combine two ingredients: (i) TNG50 of the IllustrisTNG project, a high-resolution magnetohydrodynamical galaxy formation simulation in a cosmological volume of ${\sim}50^{3}$\,cMpc$^{3}$, and (ii) a full Monte Carlo \Lya radiative transfer method to capture the impact of resonant scattering. We use the resulting \Lya predictions to explore the properties and detection rates of filamentary structures of the large-scale cosmic web in \Lya emission. Our main results are:

\begin{enumerate}
  \item Our \Lya emission model includes collisional excitations and recombinations in diffuse gas, as well as emission sourced by star forming regions. We couple this to an empirically motivated dust treatment, which we calibrate against the observed LAE luminosity function at $z=2$~(Section~\ref{sec:methodology}). Our calibrated model reasonably reproduces other related observables: the galaxy UV luminosity function, the \Lya escape fraction versus stellar mass, and the rest equivalent width distribution. It predicts a flattening \Lya luminosity versus galaxy/halo mass observable relation (Section~\ref{sec:lfs}).
  \item Within the fiducial model, radiative transfer of \Lya photons substantially alters the budget, spatial distribution, and observability within halos, filaments and voids. Collisional excitations in cold gas in the circumgalactic medium (CGM) of galaxies dominate the global emission budget. Star-formation from within galaxies plays a significant role, and recombinations can be substantial around local ionizing radiation from nearby AGN (Section \ref{sec:mechsandorigins}).
  \item Our main theoretical result is that diffuse \Lya cosmic web filaments are dominated by emission which originates from \textit{within} galaxies and their gaseous halos, and not from the intergalactic medium (IGM) itself. \Lya filaments are therefore illuminated by scattered photons which are intrinsically emitted within nearby halos. This emission arises predominantly from `intermediate mass' halos between $10^{10} - 10^{11}$\msun. The ultimate physical origin is collisional excitations in their circumgalactic media as well as due to central, young stellar populations (Sections \ref{sec:visualinspection} $-$ \ref{sec:sbdist}).
\end{enumerate}

\noindent Considering the prospects for observing \Lya filaments, we find that:

\begin{enumerate}
  \item Identifying filaments as extended regions (i.e. isophotal areas) above a surface brightness threshold of $10^{-20}$\sbunits, we predict that `large' filaments with extent $L \ge 400$\,pkpc have a space number density of $2 \times 10^{-3}$\,cMpc$^{-3}$. These \Lya structures directly trace large-scale dark matter filaments with characteristic overdensities of $\sim$\,10, and are responsible for $\sim 60\%$ of the global \Lya density $\dot{\rho}_{\mathrm{Ly}\alpha}=4.4\cdot 10^{40}$\ergs\,cMpc$^{-3}$ (Section \ref{sec:identification}).
  \item With respect to the relationship between \Lya filaments and LAEs, we find that essentially all filaments host one or more low luminosity \Lya emitters with $L > 10^{39}$\ergs. However, only large filaments tend to contain a bright $L > 10^{41}$\ergs LAE. In all cases, the diffuse regions ($<10^{-19}$\sbunits) of \Lya filaments are powered by faint LAEs due to scattered photons. The emission arising within the IGM itself is negligible. (Section \ref{sec:dicussCW}).
  \item We quantify the probability of detecting \Lya filaments for both blind and targeted survey strategies, covering instrument footprint sizes ranging from KCWI, to MUSE, BlueMUSE, and HET-VIRUS. The surface brightness levels of \Lya filaments are within reach of current integral field spectrograph instruments. However, at such moderate surface brightness thresholds, filaments are sufficiently rare that large survey areas are needed (Section~\ref{sec:prospects}).
\end{enumerate}

In the near future, new telescopes and instruments will continue to push to higher sensitivities and larger survey sizes, ultimately with the hope of imaging the large-scale cosmic web directly in $\Lya$ emission. From the theoretical side, our overall model approach and its associated \Lya photon-level output will enable us to explore several promising directions. First, we can study the efficacy of more advanced filament detection and characterization algorithms. Second, we have not yet considered the spectral (i.e.\ wavelength) dimension of these datasets, the information content therein, including comparisons of emergent $\Lya$ spectra with data.

Simultaneously, two key areas of improvement are the inclusion of intrinsic \Lya emission from AGN, and a more physically motivated dust treatment. To this end, galaxy formation simulations beyond TNG50 will enable us to push towards smaller spatial scales, permitting self-consistent dust models and explicit treatments of the escape of \Lya radiation from the star-forming ISM. Complexity is both a curse and a blessing, as the rich multi-scale nature of \Lya emission simultaneously informs the physics of emission, radiative transfer, and the underlying galaxy formation model. 

\section*{Acknowledgements}

CB and DN acknowledge funding from the Deutsche Forschungsgemeinschaft (DFG) through an Emmy Noether Research Group (grant number NE 2441/1-1). We also thank the Hector Fellow Academy for their funding support. This work was further supported by the Deutsche Forschungsgemeinschaft (DFG, German Research Foundation) under Germany's Excellence Strategy EXC 2181/1 - 390900948 (the Heidelberg STRUCTURES Excellence Cluster).
The TNG50 simulation was run with compute time granted by the Gauss Centre for Supercomputing (GCS) under Large-Scale Projects GCS-DWAR on the GCS share of the supercomputer Hazel Hen at the High Performance Computing Center Stuttgart (HLRS). GCS is the alliance of the three national supercomputing centres HLRS (Universit{\"a}t Stuttgart), JSC (Forschungszentrum J{\"u}lich), and LRZ (Bayerische Akademie der Wissenschaften), funded by the German Federal Ministry of Education and Research (BMBF) and the German State Ministries for Research of Baden-W{\"u}rttemberg (MWK), Bayern (StMWFK) and Nordrhein-Westfalen (MIWF). Additional simulations and analysis were carried out on the Vera machine of the Max Planck Institute for Astronomy (MPIA) and systems at the Max Planck Computing and Data Facility (MPCDF).

\section*{Data Availability}

Data directly related to this publication and its figures is available on request from the corresponding author. The IllustrisTNG simulations, including TNG50, are publicly available and accessible at \url{www.tng-project.org/data} \citep{Nelson19}.

\bibliographystyle{mnras}
\bibliography{references}

\label{lastpage}
\end{document}